\documentclass[
  journal=largetwo,
  manuscript=article-type,
  year=2024,
  volume=37,
]{cup-journal}

\usepackage{amsmath}
\usepackage[utf8]{inputenc}
\usepackage{aas-macros}
\usepackage{booktabs}
\usepackage{multirow,overpic}
\usepackage{upgreek}
\usepackage{graphicx}                                       
\usepackage{amssymb}                                        
\usepackage{collcell}                                       
\usepackage{float}                                          
\usepackage{flafter}                                        
\usepackage{booktabs}                                       
\usepackage{threeparttable}                                 
\usepackage{gensymb}                                        

\title{The Fast Radio Burst Population Energy Distribution}

\author{W.R.~Arcus}
\affiliation{International Centre for Radio Astronomy Research, Curtin University, GPO Box U1987, Perth, WA 6845, Australia}
\email[W.R.~Arcus, C.W.~James]{wayne.arcus@icrar.org, clancy.james@curtin.edu.au}

\author{C.W.~James}
\affiliation{International Centre for Radio Astronomy Research, Curtin University, GPO Box U1987, Perth, WA 6845, Australia}

\author{R.D.~Ekers}
\affiliation{Australia Telescope National Facility, CSIRO, Space and Astronomy, PO Box 76, Epping, NSW 1710, Australia}
    \alsoaffiliation{International Centre for Radio Astronomy Research, Curtin University, GPO Box U1987, Perth, WA 6845, Australia}

\author{J-P.~Macquart}
\affiliation{International Centre for Radio Astronomy Research, Curtin University, GPO Box U1987, Perth, WA 6845, Australia}

\author{E.M.~Sadler}
\affiliation{Sydney Institute for Astronomy, School of Physics A28, The University of Sydney, NSW 2006, Australia}
\alsoaffiliation
{Australia Telescope National Facility, CSIRO, Space and Astronomy, PO Box 76, Epping, NSW 1710, Australia}

\author{R.B.~Wayth}
\affiliation{International Centre for Radio Astronomy Research, Curtin University, GPO Box U1987, Perth, WA 6845, Australia}

\author{K.W.~Bannister}
\affiliation{Australia Telescope National Facility, CSIRO, Space and Astronomy, PO Box 76, Epping, NSW 1710, Australia}
\alsoaffiliation{Sydney Institute for Astronomy, School of Physics A28, The University of Sydney, NSW 2006, Australia}

\author{A.T.~Deller}
\affiliation{Centre for Astrophysics and Supercomputing, Swinburne University of Technology, John St, Hawthorn, VIC, 3122, Australia}

\author{C.~Flynn}
\affiliation{Centre for Astrophysics and Supercomputing, Swinburne University of Technology, John St, Hawthorn, VIC, 3122, Australia}

\author{M.~Glowacki}
\affiliation{International Centre for Radio Astronomy Research, Curtin University, GPO Box U1987, Perth, WA 6845, Australia}

\author{A.~C.~Gordon}
\affiliation{Center for Interdisciplinary Exploration and Research in Astrophysics (CIERA) and Department of Physics and Astronomy, Northwestern University, Evanston, IL 60208, USA}

\author{L. Marnoch}
\affiliation{School of Mathematical and Physical Sciences, Macquarie University, NSW 2109, Australia}
\alsoaffiliation{Astrophysics and Space Technologies Research Centre, Macquarie University, Sydney, NSW 2109, Australia}
\alsoaffiliation{Australia Telescope National Facility, CSIRO, Space and Astronomy, PO Box 76, Epping, NSW 1710, Australia}
\alsoaffiliation{ARC Centre of Excellence for All-Sky Astrophysics in 3 Dimensions (ASTRO 3D)}

\author{S.D.~Ryder}
\affiliation{School of Mathematical and Physical Sciences, Macquarie University, NSW 2109, Australia}
\affiliation{Astrophysics and Space Technologies Research Centre, Macquarie University, Sydney, NSW 2109, Australia}

\author{R.~M.~Shannon}
\affiliation{Centre for Astrophysics and Supercomputing, Swinburne University of Technology, John St, Hawthorn, VIC, 3122, Australia}


\newcommand{\dmunits}{\,pc\,cm$^{-3}$\,}                    
\newcommand{\JperHz}{\ensuremath{\mathrm {J\,Hz}^{-1}}}                     
\newcommand{\FRB}[1]{FRB\,#1}                               
\newcommand{\VMax}{$V_{\mathrm{max}}$\,}                    
\newcommand{\zmax}{$z_{\mathrm{max}}(\theta)$\,}            
\newcommand{\VonVmax}{$V/V_{\mathrm{max}}$\,}               
\newcommand{\VonVmaxbar}{$\left< V/V_{\mathrm{max}}\right>$\,}               
\newcommand{\OneonVmax}{$1/V_{\mathrm{max}}$\,}             
\newcommand{\LocalizedSample}{Localised High S/N Sample}    
\newcommand{\FullSample}{Full Sample}                       
\newcommand{\DefaultPlotSize}{0.60}                         

\newcommand{\includepic}[1]{\includegraphics[width=7.0cm, height=6.0cm, keepaspectratio=true]{#1}}
\newcolumntype{i}{@{\hspace{2ex}}>{\collectcell\includepic}c<{\endcollectcell}}

\defcitealias{MacquartEkers2018b}{\,M\&E\:2018 }

\newcommand{\shannon}{Shannon et al., in prep.}

\defcitealias{Arcusetal20}{Paper~I}

\defcitealias{Arcusetal22}{Paper~II}

\begin{document}


\begin{abstract}
We examine the energy distribution of the fast radio burst (FRB) population using a well-defined sample of 63 FRBs from the ASKAP radio telescope, 28 of which are localised to a host galaxy. We apply the luminosity-volume ($V/V_{\mathrm{max}}$) test to examine the distribution of these transient sources, accounting for cosmological and instrumental effects, and determine the energy distribution for the sampled population over the redshift range $0.01 \lesssim z \lesssim 1.02$.
We find the distribution between $10^{23}$ and $10^{26}$\,J\,Hz$^{-1}$ to be consistent with both a pure power-law with differential slope $\gamma=-1.96 \pm 0.15$, and a Schechter function with $\gamma = -1.82 \pm 0.12$ and downturn energy $E_{\rm max} \sim 6.3 \, \cdot 10^{25}$\,J\,Hz$^{-1}$. We identify systematic effects which currently limit our ability to probe the luminosity function outside this range and give a prescription for their treatment.
Finally, we find that with the current dataset, we are unable to distinguish between the evolutionary and spectral models considered in this work.
\end{abstract}



\section{Introduction}
\label{sec:Intro}
Fast Radio Bursts (FRBs) are short duration (millisecond timescale), dispersed, transient events in the radio spectrum known to originate from cosmological distances\citep{Lorimer2007,Thornton2013,Chatterjeeetal2017,Bannisteretal2019}. Current research has two major directions: to determine their progenitor source(s) and to use them as cosmological probes \citep{Macquartetal2020}. Accordingly, the FRB population statistics continues to be a topic of considerable conjecture \citep[see e.g.,][and references therein]{2022A&ARv..30....2P}.

Determining the intrinsic energy distribution (i.e., luminosity function) of FRBs has, hitherto, proven to be problematic. The first impediment stems from radio telescopes with optics that make an accurate determination of the FRB location within the telescope beam difficult, such as Parkes/Murriyang \citep[e.g.][]{Thornton2013,SUPERB1}, UTMOST \citep{Farah2019}, FAST \citep{2021ApJ...909L...8N}, and the first FRB searches with CHIME \citep{CHIME_catalog1_2021}. This complicates the construction of a fluence-complete sample and determining the effective survey area \citep[see][]{KeanePetroff2015,MacquartEkers2018a}. This issue is effectively mitigated when using telescope arrays to search for FRBs, as pioneered by the Australian Square Kilometre Array Pathfinder (ASKAP) telescope --- which uses phased array feeds (PAFs) to provide a wide field of view with dense coverage of the focal plane --- permitting reliable estimates of the survey area and FRB fluences to be made \citep{Bannisteretal2017,Shannonetal2018}.

The second impediment is the difficulty in obtaining an FRB distance estimate, which yields the FRB energy and survey volume. This requires either arcsecond-precision FRB localisations, thereby permitting the identification of the host galaxy, or the existence of a relation between the FRB dispersion measure, DM, and redshift, $z$. ASKAP has helped provide both, with a large sample of FRBs localised to their host galaxies \citep{Shannon2024}, and the establishment of a $z$-DM relationship, known as the Macquart Relation \citep[][]{Macquartetal2020}. Other instruments with similar capabilities include DSA~110 \citep{2024ApJ...967...29L}, MeerKAT \citep{2022MNRAS.514.1961R}, the VLA \citep{2018ApJS..236....8L} and CHIME's outriggers \citep{2021AJ....161...81L}.

Several authors have modelled FRB observations to determine the best-fitting FRB population parameters \citep{Luo2020,James_etal_2021_1,Shin2022,2024arXiv240804878H}. However, these fits rely on assumptions about the functional form of the FRB energy distribution and source evolution, which may differ from that of other classes of transients. A non-parametric way to determine both --- the \VonVmax\ method --- was described by \citet{Schmidt_1968}, in the context of studies of the quasar population. The simplest application of this method is to test for a spatially uniform distribution of FRB sources, which has been applied to FRB data by several authors \citep{Opperman2016,Shannonetal2018,Locatellietal19}. Others have applied the analysis determining the FRB energy distribution from non-localised FRBs \citep{Lu2019,Hashimoto2022,Li2023}, which has the aforementioned uncertainties of fluctuations in the Macquart relation.

In this work, we update these analyses using FRBs detected by the Commensal Real-time ASKAP Fast Transients \citep[CRAFT;][]{2010PASA...27..272M} survey with ASKAP \citep[][]{2021PASA...38....9H}. In particular, we use a large sample of FRBs with known redshift, allowing for the first time an accurate measurement of both $V$ and \VMax for FRBs. This allows unbiased estimates of their energy and spatial distributions to be used.

In \S \ref{sec:LuminosityFunction} we review the volumetrics and formation of the energy distribution for a sample being analysed. We then apply the approach to the ASKAP sample and outline our results and observations in \S \ref{sec:ApplicationToASKAP}. In \S \ref{sec:Discussion} we discuss the implications of the energy distributions and present our conclusions in \S \ref{sec:Conclusion}.

\section{The Energy Function}
\label{sec:LuminosityFunction}
\subsection{The $V/V_{\mathrm{max}}$ Test}

The discovery of FRBs in 2007 \citep{Lorimer_2007} has many similarities to the discovery of quasars \citep{Schmidt_1963}; both are new classes of extragalactic objects catalogued in surveys with well-defined but complex detection limits.

To estimate the spatial distribution and luminosity function of quasars (then referred to as QSOs), \citet{Schmidt_1968} introduced the \VonVmax parameter which, for each source, provides a measure of its position within the maximum volume over which it would have been observed in the complete sample. Due to the uncertainty in cosmological models at the time, \citet{Schmidt_1968} calculated volumes in co-moving coordinates using two cosmological models: luminosity distance $D_L \propto z$, and $D_L \propto z (1+0.5 z)$. Schmidt notes that \VonVmax provides a very simple test of uniformity for the spatial distribution in a sensitivity-limited sample, with an expectation value $<$\VonVmax$> = 0.5$. In the case of quasars, \VonVmax was found to be significantly larger than 0.5, and Schmidt concluded that the sample was strongly evolving. The expected uniformity in \VonVmax was achieved by weighting the Cartesian volume $V \sim D_L^3$ by an assumed source evolution of $(1+D_L)^2$.
Schmidt then estimated the local luminosity function by using \OneonVmax to weight the contribution to the spatial density from each source separately, and then grouped the sources in luminosity bins, wherein these luminosities were converted to the rest frequency.

Like quasars, FRBs are also cosmologically distributed, and the problems of analysing their redshift evolution and luminosity function are very similar. The \VonVmax method requires a complete sample of sources above a well-defined flux (or fluence) limit. Even if the redshifts of these sources are unknown or poorly defined, the mean value of \VonVmax can indicate whether the sources are distributed uniformly through the sample volume. A uniform distribution (with $<$\VonVmax$> \approx 0.5$) implies a population that is non-evolving (i.e., not changing with distance) within the sample volume, while a larger or smaller value implies either an incompleteness in selection, or a population that undergoes some form of redshift evolution within the sample volume. 

For a source population where redshift measurements are available for individual objects, and where there is also little or no evolution within the sample volume, a luminosity function can be calculated by summing the values of 1/\VMax within different luminosity bins. Local radio luminosity functions for large, complete samples have been calculated by several authors \citep[e.g.][]{condon2002,sadler2002,best2005,mauch2007}, and \cite{pracy2016} calculated the radio luminosity function for high- and low-excitation radio galaxies in several redshift bins out to $z\sim0.75$. \citet{avni1980} extended \citet{Schmidt_1968}'s method to samples with different completeness limits in two (or more) different parameters; this technique may be used to measure a bivariate luminosity function, e.g., a set of radio luminosity functions for different bins in optical luminosity \citep{mauch2007} or black hole mass determination \citep{best2005}. 

If the redshift range covered by a survey is large enough that redshift evolution occurs within the sample-volume (i.e., \VonVmax has a value significantly different from 0.5), then this evolution must be taken into account. Examples from the literature include studies of the luminosity function of gamma-ray bursts \citep{schmidt2009} and the redshift evolution of powerful radio galaxies \citep{dunlop1990}. 

Schmidt's methods may be applied directly to the FRB population, with a few significant differences. Since FRBs are transient rather than static sources of emission, the observing time should be included in the analysis as well as the survey area.
Transients are typically characterised by their fluence and energy distribution rather than their flux and luminosity function. To keep notation consistent with \citet{Schmidt_1968}, hereinafter we refer to FRB luminosities and their radio luminosity function (RLF) when describing the distribution of their spectral energy density, $E_\nu$.
If the positions are not determined well enough during the outburst, the location in the field of view cannot be determined. Thus the sensitivity of the telescope beam at the detection point, hence the correction for that sensitivity, cannot be made. For the population of FRBs that have not been observed to repeat, only surveys which determine the position in the field of view can therefore be used --- significantly reducing the applicable sample size.

First, we use the \VonVmax\ test to check whether the FRBs in our sample are uniformly distributed in space. Then, following Schmidt, we use \OneonVmax\ for each FRB to estimate its contribution to the density of FRBs of that luminosity. The estimation of \VMax\ is the critical aspect introduced by this analysis: for FRB surveys it can be applied on a per source basis, provided the survey detection limit, the detected signal-to-noise (S/N) and the position in the field of view are known for each FRB. This requirement significantly limits the sample of FRBs that can be used, and we therefore confine our analysis to suitable FRBs from the CRAFT survey, which satisfy these criteria. We do this for FRBs with known host galaxies for which $V$ and \VMax can be calculated. We also investigate the effect of using DM as a distance proxy by comparing this result to that obtained when estimating FRB distances from their DMs using the cosmological DM-$z$ (`Macquart') relation \citep{Macquartetal2020}. For simplicity, in the main body of this work, we ignore FRB spectral dependence and source evolution; however, we consider both in ~\ref{sec:spectral_source}, and show that neither have a strong influence given current data. We do not explicitly calculate the time-dependence of the survey volume, thus we cannot calculate the FRB rate. 
Moreover, since we use data from both ASKAP's Fly's Eye and Incoherent Sum (ICS) modes in different proportions for the two samples, the relative normalisation is arbitrary. We discuss this further in ~\ref{sec:VonVMaxNonUniformSensitivity}.

The ratio of volumes from which the FRB has been detected, $V$, to that in which it could have been detected, $V_{\mathrm{max}}$, is a measure of the position of the detected event within the probed volume. The statistic $\langle V/V_{\mathrm{max}}\rangle$ is the algebraic mean of events in a sample and is expressed as
\begin{equation}
    \langle V/V_{\mathrm{max}}\rangle = \frac{1}{N} \sum_{i=1}^{N} \frac{V_{i}}{V_{\mathrm{max},i}}
    \text{,}
    \label{eqn:MeanVonVmax}
\end{equation}
\noindent where $i$ represents the $i^{\mathrm{th}}$ event in a sample of $N$ events. A spatially uniform sample would be uniformly distributed over the range $\left[0, 1\right]$ with $\langle V/V_{\mathrm{max}}\rangle=0.5$ \citep{Schmidt_1968}. The luminosity function may be determined from a contribution of each event by taking the reciprocal of the volume in which each event could have been observed (i.e., $1/V_{\mathrm{max},i}$), and binning in terms of energy.

In the case of an evolving population \citep[e.g., source density evolving with redshift, or source luminosity variations; ][]{Schmidt_1968,MacquartEkers2018b} or incorrect assumptions regarding the nature of the volume, the distribution given through equation (\ref{eqn:MeanVonVmax}) will not, in general, be uniform. Re-weighting $V$ by the correct source density, $\psi$, within that volume, i.e., $V \rightarrow V^\prime = V \cdot \psi\left(V\right)$, would, however, restore the distribution to uniformity.

Measurement of the FRB luminosity distribution presents a number of complications not typically encountered with static sources, since it is not possible to find all objects by scanning an area of sky with uniform sensitivity. For a sample of static sources, one may clearly define the volume over which a source would have been detectable, viz., the volume of a spherical sector whose radius is governed by the luminosity distance out to which an object could have been detected, given the telescope sensitivity. For radio transients such as FRBs, however, this is not the case: the instantaneous sensitivity across the field of view (FoV), when the FRB is detected, is non-uniform and the volume probed is therefore not a section of a sphere. When one is interested in the event rate rather than the source density per comoving volume, the additional effect of the observing time and time dilation as a function of distance needs to be taken into account.

The spectral energy density, $E_{\nu, 0}$, of a given FRB, its observed fluence, $F_{\nu,0}$, and its luminosity distance, $D_{L}$, are related via equation (\ref{eqn:FluenceRelation})
\begin{equation}
    E_{\nu,0} = \frac{4 \pi F_{\nu,0} D_L^2}{ (1+z_b)^{2+\alpha}}
    \text{,}
    \label{eqn:FluenceRelation}
\end{equation}
\\
where $z_b$ is the redshift of the FRB and $\alpha$ the fluence spectral index. We define $\alpha:F_{\nu}\propto\nu^{\alpha}$ --- this is now common usage, however it is the opposite sign convention to that used in \citet{MacquartEkers2018b} and subsequently in \citet{Arcusetal20} and \citet{Arcusetal22}.

\begin{figure}
   \centering
   \includegraphics[scale=\DefaultPlotSize]{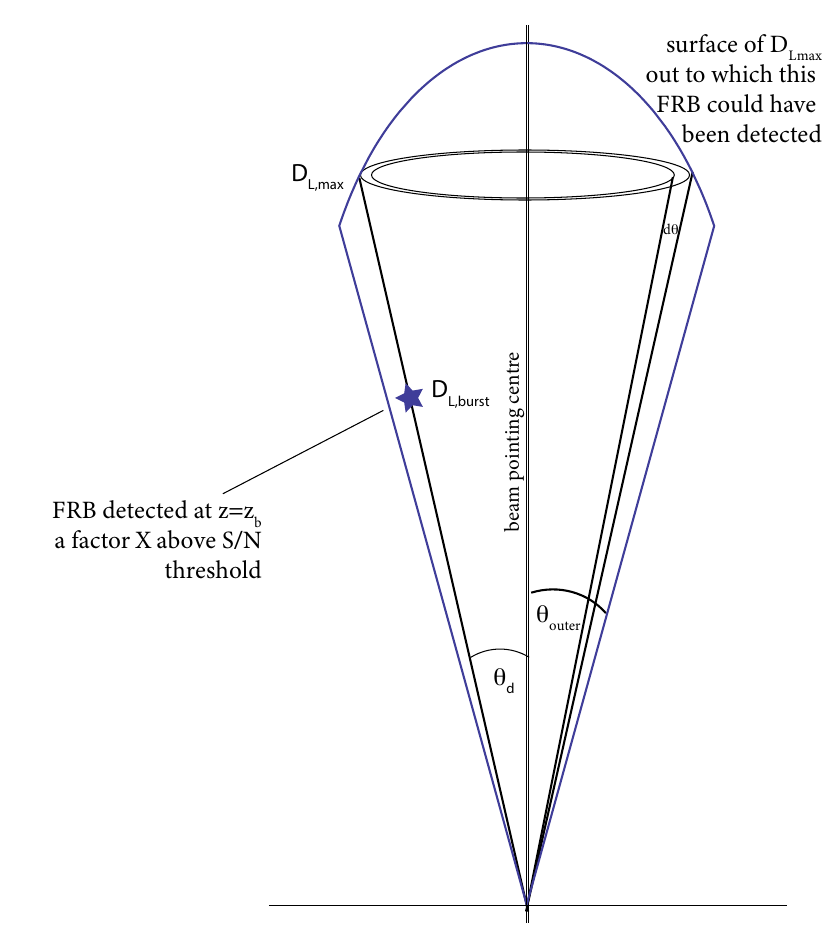}
   \caption
   {
       The geometry of the \VMax region defining the total comoving volume out to which a given FRB may be detected with a S/N being a factor of $X$ above the threshold detection S/N for a generic beam. Note that \VMax must be computed separately for each FRB since the S/N of a given FRB depends upon both the FRB fluence and duration: the $D_{L,{\mathrm{max}}}$ surface cannot be specified solely in terms of a threshold fluence.
   }
   \label{fig:Vmax}
\end{figure}

\subsection{The Survey Volume}
\label{sec:SurveyVolume}

\subsubsection{The Maximum Volume Probed by a Generic Beam}
\label{sec:VolMaxGeneric}
Consider a FRB event occurring at a given offset, $\theta_{d}$, from the beam centre of a generic telescope beam at a signal to noise (S/N) value that is a factor of $X$ above the cut-off S/N flux threshold $S_{\mathrm{cutoff}}$ (see Figure~\ref{fig:Vmax}). We would like to know over what volume this particular event, with FRB spectral energy density, $E_\nu$, and burst-width, $\Delta t$, could have been detected. If the telescope beam is circularly symmetric, the comoving volume of space probed out to a redshift $z_{\mathrm{max}}$ is given by

\begin{equation}
    \begin{split}
    V_{\mathrm{max}}
    &= \int \int \frac{D_H}{E(z)} \frac{D_L^2(z)}{(1+z)^2} dz d\Omega \\
    &=2 \pi \int_0^{\theta_{\mathrm{outer}}} \int_0^{z_{\mathrm{max}}(\theta)}
    \frac{D_H}{E(z)} \frac{D_L^2(z)}{(1+z)^2} \sin \theta d\theta dz
    \text{,}
    \end{split}
    \label{eqn:Vmax}
\end{equation}
\\

\noindent where $\Omega$ is the solid angle of the telescope beam on the sky; $\theta$, the bore-sight angle of the telescope beam; and $\theta_{\mathrm{outer}}$ the outermost detectable angle of the beam. Moreover, \zmax is the redshift of the maximum luminosity distance that an event could be detected in the telescope beam and $D_H$ and $D_L(z)$ are the Hubble distance and luminosity distance for a given redshift, $z$, respectively.

We write $z_{\mathrm{max}}(\theta)$ as an explicit function of $\theta$ to emphasise that the telescope probes to a larger redshift at the beam centre relative to its periphery. We take the integral over the angular distance to extend out to an effective beam cut-off point; the objective here being to find \zmax for a given FRB so that the effective survey volume may be determined.

We may compute the maximum detectable luminosity distance for each FRB at its particular location within the telescope beam via equation (\ref{eqn:FluenceRelation}), to determine $E_{\nu, 0}$, then find the luminosity distance at which the FRB of this energy density would be detectable at the threshold $S_{\mathrm{cutoff}}$.

An additional complication is that the detection S/N is not determined just by the FRB flux density; rather, S/N is proportional to a product involving the FRB flux density and its duration. Thus the threshold fluence is obtained by solving

\begin{equation}
    \begin{split}
    \frac{S_{\mathrm{cutoff}}}{S} \equiv X(\theta_d)
    &= \frac{S_{\nu,0} \Delta t_0^{1/2}}{S_{\nu,{\mathrm{cutoff}}} \Delta t_{\mathrm{cutoff}}^{1/2} } \\
    &= \frac{F_{\nu,0} }{F_{\nu,{\mathrm{cutoff}}}} \frac{(1+z)^{-1/2}}{(1+z_{\mathrm{max}})^{-1/2}}
    \text{.}
    \end{split}
    \label{eqn:Xequn}
\end{equation}
\\

\noindent The solution of equation (\ref{eqn:Xequn}) yields the following transcendental equation for the limiting detectable fluence for a given FRB:

\begin{equation}
    F_{\nu,{\mathrm{cutoff}}} = \frac{F_{\nu,0}}{X(\theta_d)} \left( \frac{1+z_b}{1+z_{\mathrm{max}}} \right)^{-1/2}
    \text{,}
    \label{eqn:Fcutoff}
\end{equation}
\\

\noindent and we solve this equation to determine $z_{\mathrm{max}}(\theta)$.

Yet a further complication is that the telescope detection efficiency decreases with increasing DM, which is nearly linearly proportional to redshift at $z \lesssim 1$ \citep[see e.g., ][]{Arcusetal22}. If the telescope efficiency, $\eta$, is written in terms of DM, the maximum luminosity distance out to which the FRB is detectable\footnote{We may see this by directly placing $\eta$ for the flux density terms in equation (\ref{eqn:Xequn}).} is given by

\begin{equation}
    \begin{split}
    D_{L,{\mathrm{cutoff}}} = D_{L,b} \, X^{1/2}(\theta_d) &
    \left( \frac{1+z_{\mathrm{max}}}{1+z_b} \right)^{(3/2+\alpha)/2} \\ &
    \left( \frac{\eta(\mathrm{DM}(z_{\mathrm{max}}))}{\eta(\mathrm{DM}(z_b))} \right)^{1/2}
    \text{,}
    \end{split}
    \label{eqn:Dcutoff}
\end{equation}
\\

\noindent where we note explicitly that $\textrm{DM}=\textrm{DM}(z)$.

\zmax may therefore be determined by noting that $X(\theta)$ changes according to the position in the beam. For a telescope beam whose sensitivity falls off as $B(\theta)$, telescope sensitivity changes as

\begin{equation}
    X(\theta) = X(\theta_d) \frac{B(\theta)}{B(\theta_d)}\text{.}
    \label{eqn:XofTheta}
\end{equation}
\\

Thus the limiting redshift at an angular distance, $\theta$, from the beam centre may be found by solving the following equation for \zmax via

\begin{equation}
    \begin{split}
        D_{L,{\mathrm{max}}} = D_{L,b} X^{1/2}(\theta_d) &
        \left( \frac{1+z_{\mathrm{max}}}{1+z_b} \right)^{(3/2+\alpha)/2} \\ &
        \left(\frac{B(\theta)}{B(\theta_d)} \frac{\eta(\mathrm{DM}(z_{\mathrm{max}}))}{\eta(\mathrm{DM}(z_b))} \right)^{1/2}
        \text{.}
    \end{split}
    \label{eqn:SolutionofZmax}
\end{equation}
\\

\noindent Determination of \zmax via equation (\ref{eqn:SolutionofZmax}) is fully prescribed in terms of: (i) the FRB detection angle from beam centre, $\theta_{d}$; (ii) the factor above the cut-off S/N threshold, $X(\theta_d)$; (iii) the FRB redshift, $z_b$; (iv) the beam pattern, $B(\theta)$; and (v) the telescope efficiency, $\eta(\mathrm{DM}; w)$. The maximum volume in which the FRB could have been detected, $V_{\mathrm{max}}$, may then be determined using equation (\ref{eqn:Vmax}).

\subsubsection{The Detection Volume of a FRB within a Generic Beam}
\label{sec:VolGeneric}
The volume in which a FRB is detected, $V$, for a generic beam may be determined via

\begin{equation}
    \begin{split}
    V = 2 \pi
     & \left(\int_0^{\theta_{\mathrm{max}}} \int_0^{z_{\mathrm{b}}} \frac{D_H}{E(z)} \frac{D_L^2(z)}{(1+z)^2} \sin \theta d\theta dz \right. \\
      &\left. +\,\int_{\theta_{\mathrm{max}}}^{\theta_{\mathrm{outer}}} \int_0^{z_{\mathrm{max}}(\theta)}
        \frac{D_H}{E(z)} \frac{D_L^2(z)}{(1+z)^2} \sin \theta d\theta dz\right) \text{,}
    \end{split}
    \label{eqn:Vol}
\end{equation}
\\

\noindent where we integrate the constant luminosity distance of the detected FRB out to maximum angle, $\theta_{\mathrm{max}}$, at the detection threshold (i.e., at the beam cut-off fluence, see equation (\ref{eqn:ThetaMax})), then add the residual volume out to the limit of integration, $\theta_{\mathrm{outer}}$. For an overview of determining $V/V_{\mathrm{max}}$ in relation to a non-uniform sensitivity, see \ref{sec:VonVMaxNonUniformSensitivity}.

By rearranging and relabelling equation (\ref{eqn:SolutionofZmax}), and making the substitution $X(\theta_{\mathrm{max}}) B(\theta_{\mathrm{max}}) \rightarrow B^{2}(\theta_{\mathrm{max}}) X(\theta_{\mathrm{d}}) / B(\theta_{\mathrm{d}})$ via equation (\ref{eqn:XofTheta}), $\theta_{\mathrm{max}}$ may be determined by solving

\begin{equation}
    \begin{split}
    B^{2}(\theta_{d}) \, & D^{2}_{L}(z_\mathrm{b}) \, \eta(\mathrm{DM}(z_{\mathrm{b}})) \, (1+z_{\mathrm{max}})^{2+\alpha} = \\
    & B^{2}(\theta_{\mathrm{max}}) \, D^{2}_{L}(z_{\mathrm{max}}) \, \eta(\mathrm{DM}(z_{\mathrm{max}})) \, (1+z_\mathrm{b})^{2+\alpha}
        \text{.}
    \label{eqn:ThetaMax}
    \end{split}
\end{equation}

\subsection{The Volume Probed by a Specific Beam}
\label{sec:VolSpecific}
We further adapt the treatment of \S \ref{sec:SurveyVolume} to admit telescopes with arbitrary beamshapes (later in \S \ref{sec:ApplicationToASKAP} we specifically admit the ASKAP telescope beamshape).

For a beam viewing a solid angle of sky, the inverse beamshape, $\Omega(B)$ \citep[][]{James_etal_2021_1} with beam function $B(\theta)$, the maximum volume in which a FRB may have been detected, may be recast as

\begin{equation}
    \begin{split}
    V_{\mathrm{max}} &= \int \int \frac{D_H}{E(z)} \frac{D_L^2(z)}{(1+z)^2} dz d\Omega \\
    &= \int_{0}^{1} \int_0^{z_{\mathrm{max}}(B)}
    \frac{D_H}{E(z)} \frac{D_L^2(z)}{(1+z)^2} \, \Omega(B) \, dz \, dB
    \text{.}
    \end{split}
    \label{eqn:Vmax2}
\end{equation}
\\

Likewise the volume in which the FRB was detected, for a specific beam-shape, is recast as

\begin{equation}
    \begin{split}
    V =
    & 
    \int_{0}^{B_{0}(\theta_{d})} \int_0^{z_{\mathrm{max}}(B)} \frac{D_H}{E(z)} \frac{D_L^2(z)}{(1+z)^2}  \, \Omega(B) \, dz \, dB \\
    &+ \int_{B_{0}(\theta_{d})}^{1} \int_0^{z_{\mathrm{b}}}
       \frac{D_H}{E(z)} \frac{D_L^2(z)}{(1+z)^2} \, \Omega(B) \, dz \, dB
    \text{,}
    \end{split}
    \label{eqn:Vol2}
\end{equation}
\\

\noindent where $B_{0}(\theta_{d}) = B(\theta_{d}) / X(\theta_{d})$.

In order to determine the limits of integration in equation (\ref{eqn:Vol2}), and to solve equation (\ref{eqn:SolutionofZmax}) for $z_{\mathrm{max}}(B)$, we utilise an Airy beam function as the underlying beam model where necessary \citep{Arcusetal22}.

Furthermore, consistent with \citet[][see §4.3 Numerical implementation]{James_etal_2021_1}, equations (\ref{eqn:Vmax2}) \& (\ref{eqn:Vol2}) are implemented as histogram approximations (i.e., Riemann sums), such that $\int B(\Omega) \, dB \approx \sum_{i=1}^{N_B} \Omega(B) \Delta B$ where we choose $N_{B} = 10$\: \citep{James_etal_2021_1}.

As the source evolution function for FRBs is hitherto unknown, considered hypotheses generally take on the form of some function of the star formation rate \citep[SFR; e.g.][]{MacquartEkers2018b}, or delayed with respect to star formation \citep[e.g.][]{Cao2018delayedmergers}. Current fitting methods favour source evolution consistent with the cosmic star formation rate \citep{James2022H0,Shin2022}, although this is equally consistent with a generic $(1+z)^n$ model. We show in ~\ref{sec:spectral_source} that with current data, the \VonVmax\ method cannot currently discriminate between source evolution models. Accordingly, we choose the simpler case of no source evolution and set $V^\prime = V$ as discussed in \ref{sec:spectral_source}.

\section{Application to ASKAP}
\label{sec:ApplicationToASKAP}
We consider two discrete samples from the ASKAP telescope: the full set of 63 FRBs and a subset of 28 FRBs for which an identified host galaxy with measured redshift $z$ has been obtained. We treat these two samples separately in order to determine whether the use of FRBs from the DM-only inferred redshift sample yields an energy distribution consistent with those from which redshift has been independently determined. We examine the FRB population and apply the luminosity-volume- or $V/V_{\mathrm{max}}$-test to examine the source distribution of these transient sources, accounting for cosmological and instrumental effects, in order to determine the radio luminosity function (RLF) for the sampled population. In ~\ref{sec:spectral_source}, we consider both $\alpha=0$ and $\alpha=-1.5$ \citep{Macquartetal2019}, and also a cosmological evolution of the source population. However, we find little discriminating power between the two, hence we choose, hereinafter, $\alpha=0$ and no source evolution for simplicity.

We use the formalism outlined in \S~\ref{sec:VolSpecific} to determine the volumetrics necessary to conduct the $V/V_{\mathrm{max}}$-test and apply the beamshape for the ASKAP telescope, as given by \citet[][see Figure 3 of \S 4.1 therein]{James_etal_2021_1}, via the inverse beamshape, $\Omega(B)$. We choose this approach to represent a realistic beamshape for ASKAP and to avoid complications in cases where a FRB detection occurs either in multiple beams or in an outer beam.

Table~\ref{tab:LocalizedFRBDataTable} lists the candidate localised sample of FRBs along with their relevant observational parameters applicable to our analysis. Since there is some suggestion that ASKAP Incoherent Sum (ICS) observations are incomplete in the range $\text{S/N}<14$ \citep{Shannon2024}, and we wish to ensure the localised sample has minimum bias, only those FRBs for which the S/N exceeds the threshold of $\mathrm{S/N}_{\mathrm{cutoff}}=14$ were subsequently admitted for further analysis. These are listed in Table~\ref{tab:LocalizedFRBDataTableExt} and are hereinafter identified as the \LocalizedSample, comprising 19 FRBs.

Table~\ref{tab:UnlocalizedFRBDataTable} lists the candidate full sample comprising 63 ASKAP FRBs along with their relevant observational parameters applicable to our analysis. In this sample, we include the 28 FRBs localised to their host galaxies.
This constitutes the \FullSample\ (see Table \ref{tab:UnlocalizedFRBDataTable}), where the detection threshold of $\text{S/N}_{\mathrm{cutoff}} = 9.5$ as used in the CRAFT detection pipeline, is used for all FRBs irrespective of considerations of potential bias. The derived parameters of the \FullSample\ are provided in Table~\ref{tab:UnlocalizedFRBDataTableExt}, whereby redshifts, even for FRBs with measured redshift, have been estimated from their DM budget via
\begin{eqnarray}
    \mathrm{DM}_{\mathrm{Obs}} = \mathrm{DM}_{\mathrm{MW}} + \mathrm{DM}_{\mathrm{Halo}} +
    \mathrm{DM}_{\mathrm{cosmic}} + \mathrm{DM}_{\mathrm{Host}}/(1+z)
    \text{,} \label{eqn:DMBudget}
\end{eqnarray}
where $\mathrm{DM}_{\mathrm{Obs}}$ is the observed DM of the FRB, while $\mathrm{DM_{\mathrm{MW}}}$, $\mathrm{DM}_{\mathrm{Halo}}$ and $\mathrm{DM}_{\mathrm{Host}}$ are the DM contributions due to the Milky Way disc, its halo, and the FRB host environment, respectively. We set the cosmological contribution $\mathrm{DM}_{\mathrm{cosmic}}$ to its mean, $\overline{\mathrm{DM}}(z)$, using equation (\ref{eqn:meanDM}), and assume a constant host contribution of $\mathrm{DM}_{\mathrm{Host}} = 50$\dmunits and halo contribution of $\mathrm{DM}_{\mathrm{Halo}} = 50$\dmunits consistent with \citet{Arcusetal20}. $\mathrm{DM}_{\mathrm{MW}}$ is determined via the NE2001 model of \citet{Cordes_Lazio_2003}. We note that uncertainties in these quantities can be large --- of order a factor of two for $\mathrm{DM}_{\mathrm{MW}}$ \citep{Schnitzeler2012}, and perhaps a similar uncertainty for $\mathrm{DM}_{\mathrm{Halo}}$ \citep{Prochaska2019a}. Fluctuations in $\mathrm{DM}_{\mathrm{Host}}$ are not directly measured, but are estimated to be large \citep{James2022H0}. This results in potentially large fluctuations about the Macquart relation, as evinced by FRBs with exceptionally low or high DMs for their redshifts, e.g.\ FRB 20200120E with DM\, 87.82\,pc\,cm$^{-3}$ at 3.6\,Mpc
\citep{2021ApJ...910L..18B}, and FRB 20190520B with DM $1204.7$\,pc\,cm$^{-3}$ at $z=0.241$ \citep{Niu2022}.

\begin{figure}
    \centering
    \includegraphics[width=\columnwidth]{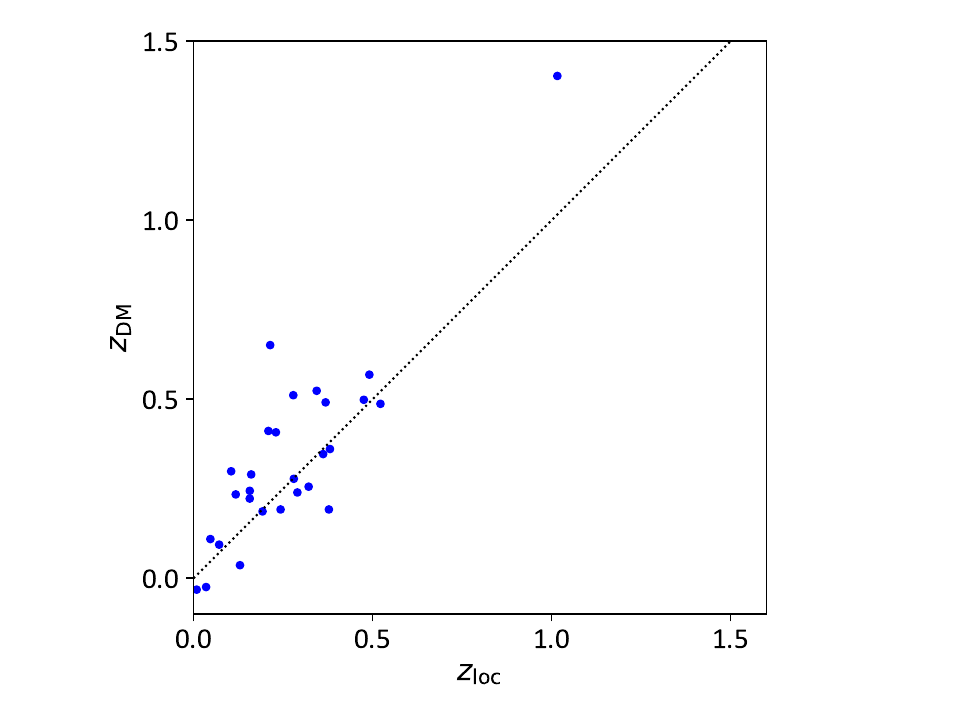}
    \caption
    {Scatter plot of spectroscopically measured host galaxy redshifts, $z_{\rm loc}$, and those derived from the Macquart relation, $z_{\rm DM}$, for the \LocalizedSample.
    }
    \label{fig:macquart_scatter}
\end{figure}

Consistent with \citet{Macquartetal2020} and \citet{Arcusetal22}, we determine the mean DM of a homogeneously distributed intergalactic medium (IGM) as given by \citet{Ioka_2003,Inoue2004}, updated to include the fraction $f_d$ of baryons in diffuse ionised gas as per \citet{2014ApJ...783L..35D}

\begin{equation}
    \begin{split}
    \overline{\mathrm{DM}}(z)=&\dfrac{3H_0c\Omega_b}{8 \pi G m_p}\\
    &\int_{0}^{z} f_d(z) \dfrac{(1+z') \left[\frac{3}{4} X_{\mathrm{e,H}}(z') + \frac{1}{8}X_{\mathrm{e,He}}(z')\right] }{\sqrt{(1+z')^3 \Omega_m + \Omega_{\Lambda}}} dz'
    \text{,}
    \end{split}
    \label{eqn:meanDM}
\end{equation}
\\
\noindent where the ionised fractions of Hydrogen and Helium are taken to be $X_{\mathrm{e,H}}=1$ for $z<8$ and $X_{\mathrm{e,He}}=1$ for $z<2.5$ respectively, or zero otherwise. Throughout this work we adopt a $\Lambda$CDM universe with the cosmological parameters $(h,\, H_{0},\, \Omega_{b},\, \Omega_{m},\, \Omega_{\Lambda},\, \Omega_{k}) = (0.7,\, 100\,h\,\mathrm{km s}^{-1} \mathrm{Mpc}^{-1},\, 0.0486,\, 0.308,\, 0.691,\, 0)$, i.e.\ an intermediate value of $H_0$ \citep{2022JHEAp..34...49A}, but otherwise in accordance with the \citet[][]{PlankResults15}. We use the estimate of $f_d(z)$ from the {\sc FRB} code base \citep{frb}. This relation between FRB redshift and expected DM was first verified by \citet{Macquartetal2020}, and is now known as the Macquart relation. Figure~\ref{fig:macquart_scatter} illustrates the scatter about the Macquart relation for the \LocalizedSample\ of FRBs. Three FRBs have an implied negative $z_{\rm DM}$, and hence are omitted from our initial analysis of the \FullSample. The effects of this are discussed in Section~\ref{sec:biases}.

\begin{figure}
    \centering
    \includegraphics[width=\columnwidth]{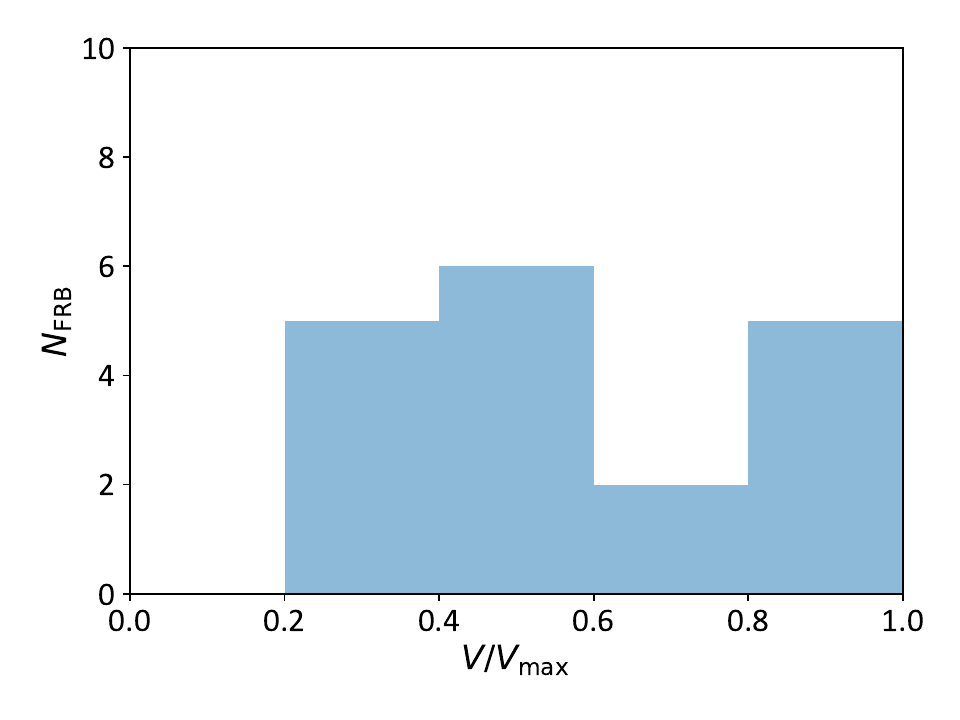} \\
    \includegraphics[width=\columnwidth]{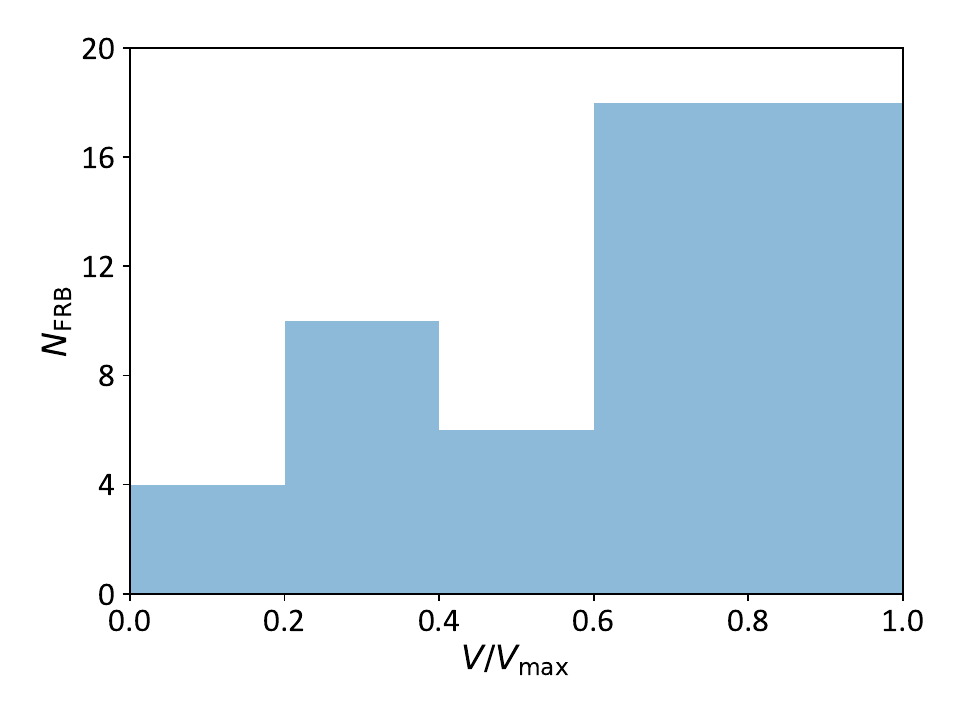}
    \caption
    {Histograms of \VonVmax\ for both the \LocalizedSample\ (top) and \FullSample\ (bottom), under the assumption of no spectral dependence ($\alpha=0$) or cosmological evolution ($n_{\rm SFR}=0$). Three FRBs with negative $z_{\rm DM}$ values have been omitted from the \FullSample.
    }
    \label{fig:VVmaxhistograms}
\end{figure}

\begin{figure*}
    \centering
    \includegraphics[width=\columnwidth]{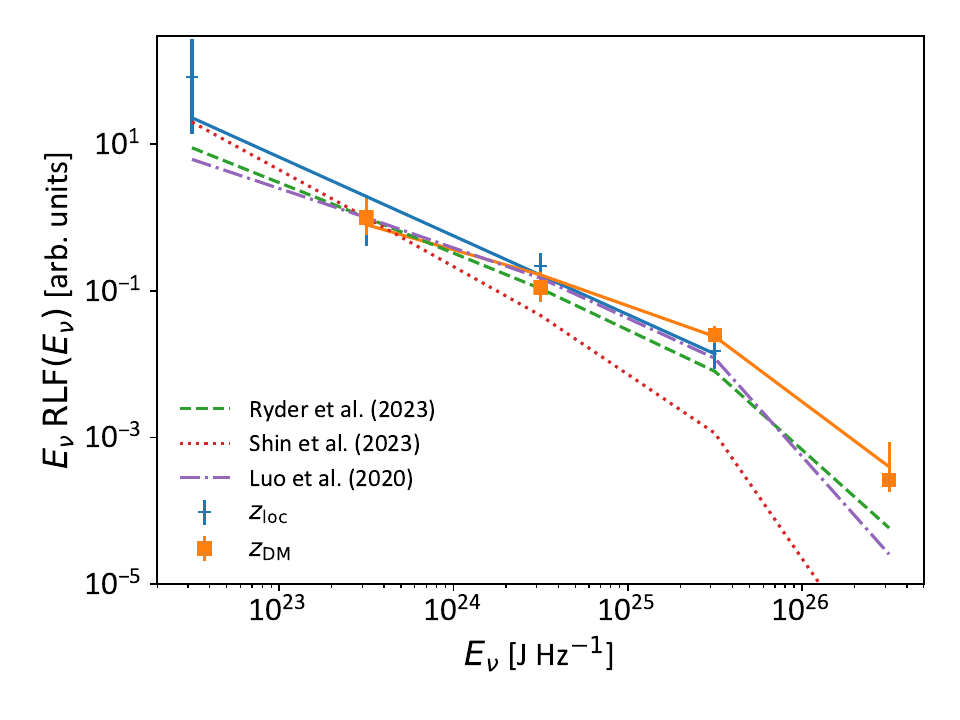}
    \caption
    {Radio luminosity functions (RLFs) calculated from  the \LocalizedSample\ (using $z_{\rm loc}$) and \FullSample\ (using $z_{\rm DM}$). The (arbitrary) normalisation is fixed to unity at the $10^{23}$--$10^{24}$ bin. The best-fit Schechter functions for each sample are depicted for reference purposes. Also shown are luminosity functions derived from ASKAP and Parkes data by \citet{Ryder2023}, CHIME data by \citet{Shin2022}, and a mixed sample by \citet{Luo2020}. The data are binned in log-space, so that the ordinate (y-axis) is effectively the RLF multiplied by the spectral fluence, $E_\nu$.
    }
    \label{fig:RLFPlots}
\end{figure*}

\section{Discussion}
\label{sec:Discussion}

\subsection{The FRB Radio Luminosity Function}


\begin{table*}
    \begin{centering}
    \begin{tabular}{lcccccc}
    \hline
    Sample & \VonVmaxbar\ & Function & $\gamma$ & $\log_{10}(E_{\rm max})$ & $\chi^2$/ndf & p-value \\
    \hline
     \FullSample\           & 0.63 & Power-law & $-1.82$ $\pm$ 0.12 & N/A & 4.24 & 0.01\\
                            & & Schechter & $-1.69$ $\pm$ 0.15 & 25.8 $\pm$ 0.39 & 2.2 & 0.34 \\ 
    De-biased \FullSample$^\dag$\, ($z = z_{\rm min}$) & 0.62 & Power-law & $-1.81$ $\pm$ 0.11 & N/A & 3.64 & 0.03 \\
                            & & Schechter & $-1.69$ $\pm$ 0.15 & 25.8 $\pm$ 0.44 & 2.13 & 0.36 \\
     \LocalizedSample\ & 0.58 & Power-law & $-2.11$ $\pm$ 0.18 & N/A & 6.15 & 0.00 \\
                            & & Schechter & $-2.11$$^\ddag$ & 40.5$^\ddag$ & 3.1 & 0.34 \\ 
    De-biased \LocalizedSample\ $z_{\rm max} = 0.7$ & 0.61 & Power-law & $-1.96$ $\pm$ 0.15 & N/A & 2.84 & 0.09 \\
                            & & Schechter & $-1.58^*$ & $25.1^*$ & ${}^*$ &  ${}^*$ \\
    \hline
    \end{tabular}
    \end{centering}\\
    \caption
    {
        Mean \VonVmax\ and best fit parameters ($\gamma$, $E_{\rm max}$) of the pure power-law and Schechter function fits to the FRB luminosity function for different data-sets, assuming no spectral dependence (i.e., $\alpha=0$) or evolution of the source population. The p-values for the power-law fits are the probability of observing a $\chi^2$ that value or higher should the power-law be the true model (high values indicate a good fit); for the Schechter function, the p-value is the probability of observing such a significant improvement in $\chi^2$ should the power-law be the true model (low values are evidence for a Schechter function).\\
     $^\dag$ Assumes FRBs with a negative $z_{\rm DM}$ are located at a distance of $z_{\rm min}=0.00024$ (2\,Mpc); fit excludes data below $10^{22}$\,J\,Hz$^{-1}$. \\
     ${}^*$ No errors can be estimated for these parameters, due to the number of degrees of freedom (ndf.) of the fit being zero.  \\
     $^\ddag$ The best-fit value of $E_{\rm max}$ is effectively infinite, rendering error calculations for the parameters meaningless. 
    \label{tab:fitparamstable}}
\end{table*}


The distributions of \VonVmax\ for both samples are shown in Figure~\ref{fig:VVmaxhistograms}. As discussed in ~\ref{sec:spectral_source}, the major deviation from uniformity is the deficit of FRBs with low \VonVmax, which cannot be rectified for any reasonable source evolution function. Hence, we proceed to calculate the radio luminosity function (RLF) from these samples, under the assumption of no spectral dependence (i.e., $\alpha=0$) and no cosmological evolution of the source population.

Figure~\ref{fig:RLFPlots} depicts the derived RLF from the \LocalizedSample\ and \FullSample. Also shown are their best-fit functions (fitted parameters given in Table~\ref{tab:fitparamstable}) and comparisons to values from the literature. A flatter RLF is preferred by the \FullSample\ ($\gamma=-1.82 \pm 0.12$) compared to the \LocalizedSample\ ($\gamma=-2.11 \pm 0.18$). At high luminosities, the \FullSample\ shows some evidence for a high-energy down-turn near $\log_{10} E_{\rm max}\,(\JperHz)=25.8 \pm 0.39$ --- likely due to the smaller \LocalizedSample\ containing no data in the $10^{26}$--$10^{27}$\,\JperHz\ bin.
Conversely, the RLF data at $E_\nu < 10^{23}$\JperHz\ from the \LocalizedSample\ shows an excess which is inconsistent with both a power-law or Schechter function, and the \FullSample\ contains no data in that luminosity bin.
Such a low-energy excess has been observed in several repeating FRBs, with low-energy peaks becoming dominant in the $\sim 10^{22}$--$10^{23}$\JperHz\ range \citep{2022RAA....22l4004N,Li2021_FAST_121102}. Furthermore, \citet{2024NatAs.tmp....5K} have found evidence for a flatter power-law index at energies above $10^{24}$\JperHz\ for FRB~ 20201124A. This suggests that apparently once-off FRBs localised with ASKAP exhibit a qualitatively similar hardening of the RLF above $10^{23}$\JperHz, though this is an ensemble average over the behaviour of many objects, and there are quantitative differences both within and between the RLFs measured for repeating FRBs; these samples may be subject to systematic biases, as discussed below.

\subsection{Systematic Biases -- \FullSample}
\label{sec:biases}
The \FullSample\ includes three low-DM FRBs with implied negative redshifts, which cannot therefore be trivially included in calculations. This results in the RLF that uses $z_{\rm DM}$ missing these events, which invariably occur in the nearby Universe, where under-fluctuations in ${\rm DM}_{\rm Host}$, ${\rm DM}_{\rm Halo}$, and/or ${\rm DM}_{\rm MW}$ could result in low measured values of ${\rm DM}_{\rm Obs}$, such that only a negative value of $z$ will satisfy equation (\ref{eqn:DMBudget}). This effect can be seen most clearly in the missing data point for the \FullSample\ in the $10^{22}$--$10^{23}$\,J\,Hz$^{-1}$ bin in Figure~\ref{fig:RLFPlots}, which in the \LocalizedSample, is entirely due to \FRB{20171020A}. One method of avoiding such a bias is to marginalise over distributions of Milky Way and host galaxy DM contributions, as performed by \citet{Locatellietal19} --- see Section~\ref{sec:locatelli} for further discussion of this approach.

The effect of this bias can be estimated by placing robust bounds on the true distance to these $z_{\rm DM}<0$\,FRBs. A lower bound assumes they are not located in Local Group galaxies, limiting the luminosity distance $D_L \gtrsim 2$\,Mpc (which equates to $z_{\rm min} = 0.00024$, ignoring peculiar velocities). An upper bound assumes that the entire DM contribution is cosmological in nature, i.e.,\ $z_{\rm max} = z_{\rm DM}({\rm DM}_{\rm cosmic} = {\rm DM}_{\rm Obs})$. We vary between these extremes, using $z = z_{\rm min} + k (z_{\rm max}-z_{\rm min})$, for $k=0,0.1,1.0$. We find that for $k\ge0.2$, the effect on the RLF is negligible. However, for very low values of $k$, the RLF extends to very low luminosities, with a dependence $\propto E_\nu^{-1.5}$, since these FRBs invariably occupy the local Universe with approximately Euclidean geometry. The case of $k=0$ only is shown in Figure~\ref{fig:RLFPlots_unbiased}.

When assuming very nearby FRBs, the low-luminosity form of the RLF is significantly changed, and we are unable to obtain consistent fits. Excluding data below $10^{22}$\,\JperHz\ produces almost identical values for $\gamma$ and $E_{\rm max}$ as those previously found for the \FullSample. We therefore conclude that this bias limits our ability to probe the low-luminosity end of the RLF.

\subsection{Systematic Biases - \LocalizedSample}
\label{sec:sys_localized}

The inclusion of \FRB{20171020A} in the \LocalizedSample\ highlights our second source of systematic bias. \FRB{20171020A} only has a confident redshift precisely because it is nearby, thus its host galaxy can be identified despite the relatively large localisation errors of the CRAFT Fly's Eye observations. The analysis presented here has no means of accounting for the likely more-distant, higher-DM FRBs of the Fly's Eye sample (those from \FRB{20170107A} to \FRB{20180525}A) which cannot be included in the \LocalizedSample. A similar effect also occurs for high-redshift --- and necessarily high-luminosity --- FRBs, the host galaxies of which may be unidentifiable due to their large distance. An example of this is \FRB{20210912A}, where optical limits on the as-yet unseen host galaxy suggests $z>0.7$, with $z=1$ implying $E_\nu = 9.7 \cdot 10^{25}\,{\rm J}\,{\rm Hz}^{-1}$ in the case of $\alpha=0$ \citep{Marnoch2023}. However, without this firm localisation, this undoubtedly energetic FRB cannot be included in the \LocalizedSample.

The biases mentioned above can be overcome in the case of the \LocalizedSample\ by using a limiting redshift $z_{\rm lim}$ such that all FRBs with $z<z_{\rm lim}$ are guaranteed to have their host galaxies identified. To do this, we first remove \FRB{20171020A} from the sample, since $z_{\rm lim}$ for the CRAFT Fly's Eye observations are poorly defined, and set $z_{\rm lim}=0.7$ for the remaining FRBs localised with ICS observations. All integrals over $z$ in the calculations for $V$ and \VMax\ in Section~\ref{sec:ApplicationToASKAP} are then terminated at $z_{\rm lim}$, while FRBs located outside this volume are excluded. Thus, the definition of \VMax\ becomes ``the volume within which this FRB would have been included in the analysis''. A limiting case of this method is to use only FRBs in a thin slice of redshift, between $z$ and $z+dz$. In such a case, $V=V_{\rm max}$, and is constant for each and every FRB, such that every FRB has equal weight in the calculation of the luminosity function, consistent with expectation.

\begin{figure}
    \centering
    \includegraphics[width=\columnwidth]{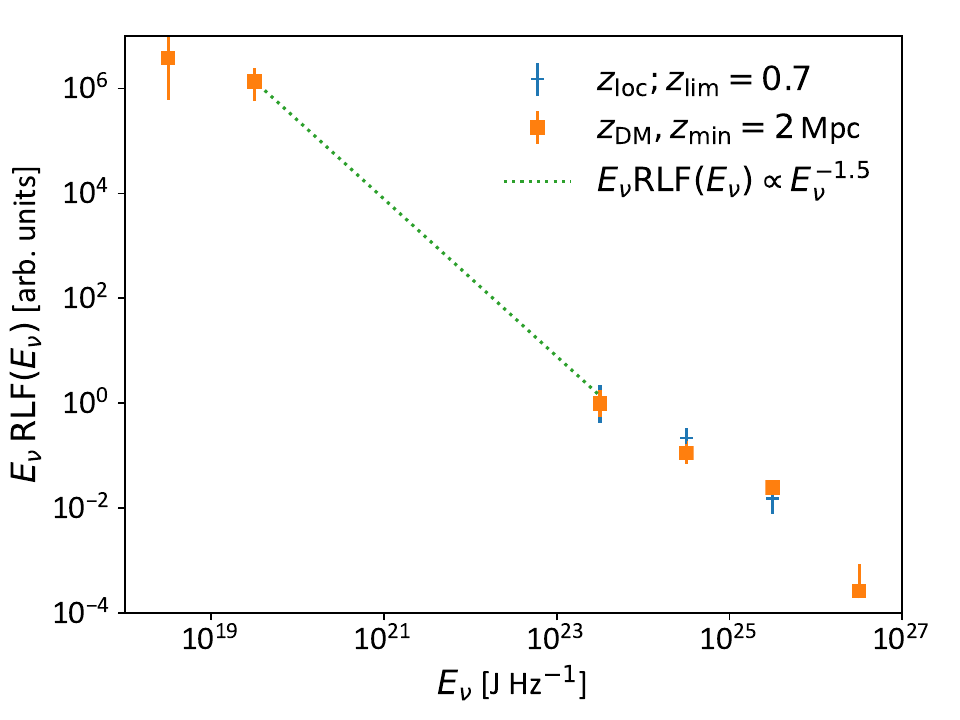}
    \caption
    {Data on radio luminosity functions (RLFs) calculated so as to account for observational biases, showing the full range allowing for a minimum distance at $z_{\rm min}$.
    The line indicating the $E_\nu$ RLF $ \propto E_\nu^{-1.5}$ is to guide the eye only.
    }
    \label{fig:RLFPlots_unbiased}
\end{figure}

Figure~\ref{fig:RLFPlots_unbiased} shows the RLF for this updated sample of FRBs. It is only measured in the range $10^{23}-10^{25}$\,J\,Hz$^{-1}$, and in this range, is consistent with a pure power-law with slope $\gamma = -1.96 \pm 0.15$ (p-value of linear fit 0.09); a Schechter function produces a much flatter differential slope $\gamma = -1.58$ and turn-over energy of $E_{\rm max} = 1.2\,\cdot 10^{25}$\,J\,Hz$^{-1}$ (errors cannot be estimated since the number of variables equals the number of points, i.e., it is statistically ill-posed).

\subsection{Comparison With Other Results}
\label{sec:comparison}

Fits of the FRB RLF have been undertaken by several authors. While many assume a 1:1 $z$--DM relation, we concentrate on those which have fully modelled uncertainties in FRB redshift given DM, and/or used a sample of localised FRBs, while accounting for selection effects as per \citet{Connor2019}. \citet{Luo2020} uses a mixed sample of mostly unlocalised FRBs from several instruments --- including Parkes and ASKAP --- to fit a Schechter function with differential index $\gamma = -1.79^{+0.31}_{-0.35}$ and $E_{\rm max} = 2.9_{-1.7}^{+11.9} \cdot 10^{25}$\JperHz\ (assuming a 1\,ms burst width and 1\,GHz bandwidth). \citet{James2022H0} uses a sample of 16 FRBs with host redshifts, and approximately 60 without, to find an index of $-1.95^{+0.18}_{-0.15}$, with \citet{Ryder2023} updating $E_{\rm max}$ to $5_{-2}^{+3} \cdot 10^{25}$\JperHz. \cite{Shin2022} fits the dispersion measure of 536 FRBs observed by CHIME to find an index of $-2.3^{+0.7}_{-0.4}$ and $E_{\rm max} = 2.38^{+5.65}_{-1.64} \cdot 10^{25}$\JperHz. These results are broadly consistent with the range of RLFs derived in this work, although they would have difficulty fitting the possible low-energy excess observed in the potentially biased \LocalizedSample, and the fit on \citet{Shin2022} has a downturn which is stronger than allowed by our data. We note that \citet{James2022H0} accounts for the biases discussed in Section~\ref{sec:biases} by not using the localisation of \FRB{20171020A} or FRBs above DM$_{\rm EG}$ of 1000\,\dmunits, while \citet{Shin2022} fits FRBs at lower frequency which may have a different underlying RLF. We therefore conclude that, given uncertainties in FRB spectral behaviour and source evolution, and possible biased effects in our own analysis, we cannot discriminate between these previous fits.

\subsection{Comparison With \citet{Locatellietal19}}
\label{sec:locatelli}

\citet{Locatellietal19} also apply the \VonVmax\ test to FRBs by comparing 23 FRBs discovered by ASKAP with 20 of the FRBs found by Parkes up to 2019. Their paper focuses on the use of the \VonVmax\ distribution to explore evolution, while our paper uses the \VMax\ method to estimate the luminosity distribution. These are different uses, and analysis of \VonVmax\ needs complete unbiased samples which is quite problematic for FRBs as discussed in Section 4.3. They find evidence for cosmological source evolution in the ASKAP sample, with \VonVmaxbar$= 0.68 \pm 0.05$, but less so for the Parkes data, with \VonVmaxbar$ = 0.54 \pm 0.04$, assuming a spectral evolution of $\alpha=1.6$.

For ASKAP, \citet{Locatellietal19} only analyse the ICS sample, which was all that was available at the time. They also include the Parkes FRB sample; however, for these FRBs the location in the beam is not known, so neither \VMax\ nor the actual beam-corrected fluences are known; we therefore excluded the Parkes FRB sample in our analysis.  This difficulty will also apply to the much larger CHIME sample. The authors also omit discussion of beam areas for Parkes and ASKAP.

As discussed in Section~\ref{sec:biases}, \citet{Locatellietal19} builds appropriate probability functions to estimate the redshift PDF($z$) instead of using a unique value. This treatment is an elegant way to avoid the bias due to negative apparent redshifts encountered when only the mean DM correction is used. We note that the {\sc zDM} code is able to produce such PDFs, e.g.\ as per \citet{2023PASA...40...29L}.

Our results for \VonVmaxbar\ for the \LocalizedSample\ and \FullSample\ are 0.58 and 0.63 respectively (0.61 and 0.62 when debiased); when we include a spectral dependence of $\alpha=-1.5$ (see ~\ref{sec:spectral_source}), we find \VonVmaxbar$=0.67$ for the \FullSample, consistent with the result of \citet{Locatellietal19}.

\begin{table*}
    \centering
    \begin{tabular}{lrrrrrrrrrl}
    \hline
    \multicolumn{1}{c}{Name} & \multicolumn{1}{c}{S/N} & \multicolumn{1}{c}{DM} & \multicolumn{1}{c}{DM$_{\mathrm{Gal}}$} & \multicolumn{1}{c}{$\nu_{c}$} & \multicolumn{1}{c}{$\Delta t$} & \multicolumn{1}{c}{$w$} & \multicolumn{1}{c}{$\theta_{d}$} & \multicolumn{1}{c}{$F_{\nu}$} & \multicolumn{1}{c}{$z_{\mathrm{loc}}$} & Reference\\
    \multicolumn{1}{c}{} & \multicolumn{1}{c}{} & \multicolumn{1}{c}{(pc cm$^{-3}$)} & \multicolumn{1}{c}{(pc cm$^{-3}$)} & \multicolumn{1}{c}{(MHz)} & \multicolumn{1}{c}{(ms)} & \multicolumn{1}{c}{(ms)} & \multicolumn{1}{c}{(\degree)} & \multicolumn{1}{c}{(Jy ms)} & \multicolumn{1}{c}{} & \multicolumn{1}{c}{}\\
    \hline
    \hline
    \FRB{20171020A}        &  19.5  & 114.1   & 38.4  & 1297.5    & 1.260 & 1.70  & 0.722  & 200   &   0.00867 & c \\
    \FRB{20180924B}        &  21.1  & 362.4   & 40.5  & 1297.5    & 0.864 & 1.76  & 0.23  & 18    &   0.3214 & h                      \\
    \FRB{20181112A}        &  19.3  & 589.0   & 40.2  & 1297.5    & 0.864 & 2.10  & 0.31  & 28    &   0.4755 & i                      \\
    \FRB{20190102C}        &  14.0  & 364.5   & 57.3  & 1271.5    & 0.864 & 1.70  & 0.23  & 16    &   0.29 & j                      \\
    \FRB{20190608B}        &  16.1  & 339.5   & 37.2  & 1271.5    & 1.728 & 6.00  & 0.36  & 28    &   0.1178 & j                      \\
    \FRB{20190611B}(*)     &  9.5   & 322.2   & 57.6  & 1271.5    & 1.728 & 2.00  & 0.48  & 10    &   0.378 & j                      \\
    \FRB{20190711A}        &  23.8  & 594.6   & 56.6  & 1271.5    & 1.728 & 6.50  & 0.08  & 36    &   0.522 & j             \\
    \FRB{20190714A}(*)     &  10.7  & 504.7   & 38.5  & 1271.5    & 1.728 & 2.90  & 0.46  & 12    &   0.2365 & k            \\
    \FRB{20191001A}        &  37.1  & 506.9   & 44.2  & 919.5     & 1.728 & 4.20  & 0.58  & 120   &   0.23 & l             \\
    \FRB{20191228A}        &  22.9  & 297.5   & 32.9  & 1271.5    & 1.728 & 2.30  & 0.49  & 67    &   0.240 & m            \\
    \FRB{20200430A}(*)     &  13.9  & 380.1   & 27.0  & 864.5     & 1.728 & 6.50  & 0.48  & 35    &   0.161 & k         \\
    \FRB{20200906A}(*)     &  10.5  & 577.8   & 35.9  & 864.5     & 1.728 & 6.00  & 0.65  & 53    &   0.36879 & m            \\
    \FRB{20210117A}        &  27.1  & 730.0   & 34.4  & 1271.5    & 1.182 & 3.2   & 0.464 & 36    &   0.214 & n            \\
    \FRB{20210320C}        &  15.3  & 384.0   & 42    & 864.5     & 1.182 & 5.4   & 1.12  & 59    &   0.28    & o              \\
    \FRB{20210807D}        &  47.1  & 251.9   & 121.2 & 920.5     & 1.182 & 10.00 & 0.453  & 100   &   0.12969 & p          \\
    \FRB{20211127I}        &  37.9  & 234.8   & 42.5  & 1271.5    & 1.182 & 1.41  & 0.16  & 35    &   0.046946 & p          \\
    \FRB{20211203C}        &  14.2  & 636.2   & 63.4  & 920.5     & 1.182 & 9.60  & 0.212  & 30    &   0.34386 & o,q          \\
    \FRB{20211212A}(*)     &  12.8  & 206.0   & 27.1  & 1632.5    & 1.182 & 2.70  & 0.77  & 131   &   0.0715 & p        \\
    \FRB{20220105A}(*)     &  9.8   & 583.0   & 22.0  & 1632.5    & 1.182 & 2.00  & 0.443  & 19    &   0.2785 & o,q          \\
    \FRB{20220501C}        &  16.1  & 449.5   & 30.6  & 863.5     & 1.182 & 6.50  & 0.516 & 32    &   0.381 & o,q          \\
    \FRB{20220610A}        &  29.8  & 1458.1  & 31.0  & 1271.5    & 1.182 & 5.60  & 0.073 & 47    &   1.016 & r,s            \\
    \FRB{20220725A}(*)     &  12.7  & 290.4   & 30.7  & 920.5     & 1.182 & 4.10  & 1.272 & 72    &   0.1926 & o          \\
    \FRB{20220918A}         &  26.4  & 657     & 41    & 1271.5    & 1.182 & 7.1   & 0.457 & 55    &   0.491   & o,q   \\
    \FRB{20230526A}         &  22.1  & 361.4   & 50    & 1271.5    & 1.182 & 4.7   & 0.383 & 34    &   0.157   & o   \\
    \FRB{20230708A}         &  31.5  & 411.5   & 50.2  & 920.5     & 1.182 & 23.6  & 0.657 & 111   &    0.105  & o   \\
    \FRB{20230718A}(*)         &  10.9  & 477     & 395.6 & 1271.5    & 1.182 & 3.5   & 0.317 & 14    &   0.035   & o   \\
    \FRB{20230902A}(*)        &  11.8  & 440     & 34.3  & 831.5     & 1.182 & 5.9   & 0.63  & 23    &   0.3619  & o   \\
    \FRB{20231226A}         &  36.7  & 329.9   & 38    & 863.5     & 1.182 & 11.8  & 0.739 & 78    &   0.1569  & o   \\
    \hline
    \end{tabular} \\
    \caption
    {
        Properties of the 28 localised ASKAP candidate FRBs for which a host galaxy redshift has been determined. FRBs identified with an asterisk below (*) are excluded from subsequent analysis since their detected $\text{S/N} < \text{S/N}_{\mathrm{cutoff}} (= 14)$.
        Variables listed in this table are:
           (i) DM - Observed DM;
           (ii) $\nu_{c}$ - Centre Frequency;
           (iii) S/N - Primary S/N;
           (iv) S/N$_{\mathrm{cuttoff}}$ - S/N threshold;
           (v) DM$_{\mathrm{Gal}}$ - DM of Milky Way disc using the NE2001 model;
           (vi) $\Delta t$ - Sample Interval;
           (vii) $w$ - Fitted Pulse-width (FWHM);
           (viii) $\theta_{d}$ - Detection Angle;
           (ix) $F_{\nu}$ - Corrected Fluence; \&
           (x) $z_{\mathrm{loc}}$ - Localized host redshift.
References are:
    a: \protect\cite{Bannisteretal2017}, b: \protect\cite{Shannonetal2018},
           c: \citet{mahony_etal_2018},
       d: \protect\cite{Macquartetal2019}, e: \protect\cite{Agarwaletal2019}, f: \protect\cite{Qiuetal19},
       g: \protect\cite{Bhandarietal2019}, h: \protect\cite{Bannisteretal2019},
       i: \protect\cite{Prochaskaetal2019}, j: \protect\cite{Macquartetal2020},
       k: \protect\cite{Heintz2020}, l: \protect\cite{Bhandari2020b},
       m: \protect\cite{Bhandari2022},  n: \protect\cite{Bhandari2023}
       o: \protect\citep{Shannon2024},
       p: \protect\cite{James_etal_2022},
       q: \protect\cite{Baptista2023},
       r: \protect\cite{Ryder2023},
       s: \protect\cite{2024ApJ...963L..34G}.
    \label{tab:LocalizedFRBDataTable}}
\end{table*}

\begin{table*}
    \centering
    \begin{tabular}{lr|rrrrr}
    \hline
    \multicolumn{1}{c}{Name} & \multicolumn{1}{c}{$D_{L}$} & \multicolumn{1}{|c}{$E_{\nu}$} &
    \multicolumn{1}{c}{$z_{\max}$} & \multicolumn{1}{c}{$D_{L,\max}$}&
    \multicolumn{1}{c}{$V_{\mathrm{\max}}$} &
    \multicolumn{1}{c}{$\mathrm{V}/V_{\mathrm{\max}}$} \\
    \multicolumn{1}{c}{} & \multicolumn{1}{c}{(Gpc)} & \multicolumn{1}{|c}{(J Hz$^{-1}$)} &
    \multicolumn{1}{c}{} & \multicolumn{1}{c}{(Gpc)}& \multicolumn{1}{c}{(Gpc$^{3}$)} & \multicolumn{1}{c}{}  \\
    \hline
    \hline
    \FRB{20171020A} &  0.037 & $3.30\cdot 10^{22}$ &  0.017 & $1.93\cdot 10^{2}$ & $6.66\cdot 10^{2}$ &  0.290  \\
    \FRB{20180924A} &  1.68 & $3.56\cdot 10^{24}$ &  0.410 & $5.09\cdot 10^{6}$ & $6.38\cdot 10^{6}$ &  0.797   \\
    \FRB{20181112B} &  2.66 & $1.09\cdot 10^{25}$ &  0.605 & $1.36\cdot 10^{7}$ & $1.66\cdot 10^{7}$ &  0.821   \\
    \FRB{20190102A} &  1.49 & $2.56\cdot 10^{24}$ &  0.303 & $2.92\cdot 10^{6}$ & $2.94\cdot 10^{6}$ &  0.995   \\
    \FRB{20190608C} &   0.549 & $8.07\cdot 10^{23}$ &  0.143 & $3.0\cdot 10^{5}$ & $3.51\cdot 10^{5}$ &  0.854  \\
    \FRB{20190711A} &  2.98 & $1.65\cdot 10^{25}$ &  0.730 & $1.64\cdot 10^{7}$ & $2.22\cdot 10^{7}$ &  0.742   \\
    \FRB{20191001A} &  1.15 & $1.25\cdot 10^{25}$ &  0.429 & $2.67\cdot 10^{6}$ & $7.14\cdot 10^{6}$ &  0.374   \\
    \FRB{20191228A} &  1.22 & $7.71\cdot 10^{24}$ &  0.389 & $2.78\cdot 10^{6}$ & $5.39\cdot 10^{6}$ &  0.516   \\
    \FRB{20210117A} &  1.06 & $3.26\cdot 10^{24}$ &  0.362 & $2.06\cdot 10^{6}$ & $4.52\cdot 10^{6}$ &  0.455   \\
    \FRB{20210320A} &  1.43 & $8.84\cdot 10^{24}$ &  0.477 & $4.36\cdot 10^{6}$ & $9.47\cdot 10^{6}$ &  0.460   \\
    \FRB{20210807D} &   0.609 & $3.47\cdot 10^{24}$ &  0.275 & $5.57\cdot 10^{5}$ & $2.10\cdot 10^{6}$ &  0.265 \\
    \FRB{20211127I} &   0.208 & $1.66\cdot 10^{23}$ &  0.079 & $2.71\cdot 10^{4}$ & $6.49\cdot 10^{4}$ &  0.418 \\
    \FRB{20211203C} &  1.82 & $6.56\cdot 10^{24}$ &  0.354 & $4.13\cdot 10^{6}$ & $4.14\cdot 10^{6}$ &  0.998   \\
    \FRB{20220501C} &  2.05 & $8.41\cdot 10^{24}$ &  0.452 & $7.35\cdot 10^{6}$ & $8.20\cdot 10^{6}$ &  0.897   \\
    \FRB{20220610A} &  6.70 & $6.22\cdot 10^{25}$ &  1.590 & $7.59\cdot 10^{7}$ & $1.07\cdot 10^{8}$ &  0.710   \\
    \FRB{20220918A} &  2.76 & $2.26\cdot 10^{25}$ &  0.927 & $1.67\cdot 10^{7}$ & $3.67\cdot 10^{7}$ &  0.455   \\
    \FRB{20230526A} &   0.749 & $1.71\cdot 10^{24}$ &  0.231 & $7.84\cdot 10^{5}$ & $1.31\cdot 10^{6}$ &  0.598 \\
    \FRB{20230708A} &   0.485 & $2.56\cdot 10^{24}$ &  0.201 & $2.92\cdot 10^{5}$ & $8.95\cdot 10^{5}$ &  0.326 \\
    \FRB{20231226A} &   0.749 & $3.91\cdot 10^{24}$ &  0.336 & $9.52\cdot 10^{5}$ & $3.56\cdot 10^{6}$ &  0.267 \\
     \hline
    \end{tabular}
    \caption
    {
        Derived properties of the 19 \textit{\LocalizedSample}\ of ASKAP FRBs for which the S/N exceeds the threshold $\text{S/N} \ge \text{S/N}_{\mathrm{cutoff}} (= 14)$, for fluence spectral indices of $\alpha=0.0$, and no source evolution. Note that this sample is a subset of the FRBs listed in Table \ref{tab:LocalizedFRBDataTable}.
    \label{tab:LocalizedFRBDataTableExt}
    }
\end{table*}

\begin{table*}
    \centering
     \caption
    {Properties of the \textit{\FullSample}. 
     Columns and references are the same as in Table~\ref{tab:LocalizedFRBDataTable}, excepting $z_{\mathrm{DM}}$ - Redshift inferred from the $z-\text{DM}$ relation.
    \label{tab:UnlocalizedFRBDataTable}}
    \begin{tabular}{lrrrrrrrrrcl}
    \hline
    \multicolumn{1}{c}{Name} & \multicolumn{1}{c}{DM} & \multicolumn{1}{c}{$\nu_{c}$} & \multicolumn{1}{c}{S/N}  & \multicolumn{1}{c}{DM$_{\mathrm{Gal}}$} & \multicolumn{1}{c}{$\Delta t$} & \multicolumn{1}{c}{$w$} & \multicolumn{1}{c}{$\theta_{d}$} & \multicolumn{1}{c}{$F_{\nu}$} & \multicolumn{1}{c}{$z_{\mathrm{DM}}$} & \multicolumn{1}{l}{Reference} \\
    \multicolumn{1}{c}{} & \multicolumn{1}{c}{(pc cm$^{-3}$)} & \multicolumn{1}{c}{(MHz)} & \multicolumn{1}{c}{} & \multicolumn{1}{c}{} & \multicolumn{1}{c}{(pc cm$^{-3}$)} & \multicolumn{1}{c}{(ms)} & \multicolumn{1}{c}{(ms)} & \multicolumn{1}{c}{(\degree)} & \multicolumn{1}{c}{(Jy ms)} & \multicolumn{1}{c}{} & \multicolumn{1}{l}{}\\
    \hline
    \hline
    \FRB{20170107A}        & 610    & 1320.5         &  16     &  37    & 1.26      & 2.4       &  0.163    & 58    & 0.506     & a \\
    \FRB{20170416A}        & 523    & 1320.5         &  13.1   &  40    & 1.26      & 5         &  0.332    & 96    & 0.413     & b  \\
    \FRB{20170428A}        & 992    & 1320.5         &  10.5   &  40    & 1.26      & 4.4       &  0.041    & 34    & 0.891     & b  \\
    \FRB{20170712A}        & 313    & 1296.5         &  12.7   &  39    & 1.26      & 3.5       &  0.281    & 52    & 0.193     & b  \\
    \FRB{20170707A}        & 235    & 1296.5         &  9.5    &  36    & 1.26      & 1.4       &  0.133    & 54    & 0.111     & b  \\
    \FRB{20170906A}        & 390    & 1296.5         &  17     &  39    & 1.26      & 2.5       &  0.237    & 74    & 0.275     & b  \\
    \FRB{20171003A}        & 463    & 1297.5         &  13.8   &  41    & 1.26      & 2         &  0.342    & 82    & 0.350     & b  \\
    \FRB{20171004A}        & 304    & 1297.5         &  10.9   &  39    & 1.26      & 2         &  0.203    & 44    & 0.184     & b  \\
    \FRB{20171019A}        & 461    & 1297.5         &  23.4   &  37    & 1.26      & 5.4       &  0.379    & 219   & 0.352     & b  \\
    \FRB{20171020A}(**)    & 114    & 1297.5         &  19.5   & 38    & 1.26      & 1.7       &  0.630    & 200   & -0.02    & b,c \\
    \FRB{20171116A}        & 619    & 1297.5         &  11.8   &  36    & 1.26      & 3.2       &  0.346    & 64    & 0.516     & b  \\
    \FRB{20171213A}        & 159    & 1297.5         &  25.1   &  37    & 1.26      & 1.5       &  0.513    & 133   & 0.024     & b  \\
    \FRB{20171216A}$^\dag$        & 203    & 1297.5         &  8      & 37    & 1.26      & 1.9       &  0.491    & 40    & 0.074     & b  \\
    \FRB{20180110A}        & 716    & 1297.5         &  35.6   &  39    & 1.26      & 7.88      &  0.430    & 422   & 0.612     & b  \\
    \FRB{20180119A}        & 403    & 1297.5         &  15.9   &  36    & 1.26      & 2.7       &  0.298    & 110   & 0.292     & b  \\
    \FRB{20180128A}        & 441    & 1297.5         &  12.4   &  32    & 1.26      & 2.9       &  0.158    & 51    & 0.336     & b  \\
    \FRB{20180128B}        & 496    & 1297.5         &  9.6    &  41    & 1.26      & 2.3       &  0.396    & 66    & 0.384     & b  \\
    \FRB{20180130A}        & 344    & 1297.5         &  10.3   &  39    & 1.26      & 4.1       &  0.380    & 95    & 0.227     & b  \\
    \FRB{20180131A}        & 658    & 1297.5         &  13.8   &  40    & 1.26      & 4.5       &  0.391    & 100   & 0.552     & b  \\
    \FRB{20180212A}        & 168    & 1297.5         &  18.3   &  31    & 1.26      & 1.81      &  0.391    & 96    & 0.041     & b  \\
    \FRB{20180315A}        & 479    & 1297.5         &  10.5   &  101   & 1.26      & 2.4       &  0.395    & 56    & 0.304     & d  \\
    \FRB{20180324A}        & 431    & 1297.5         &  9.8    &  64    & 1.26      & 4.3       &  0.494    & 71    & 0.292     & d  \\
    \FRB{20180417A}        & 475    & 1297.5         &  17.5   &  26    & 1.26      & 2.3       &  0.458    & 49    & 0.378     & e  \\
    \FRB{20180430A}(**)    & 264    & 1297.5         &  28.2   &  169   & 1.26      & 1.2       &  0.392    & 177   & -0.00     & f  \\
    \FRB{20180515A}        & 355    & 1297.5         &  12.1   &  33    & 1.26      & 1.9       &  0.191    & 46    & 0.245     & g  \\
    \FRB{20180525A}        & 388    & 1297.5         &  27.4   &  31    & 1.26      & 3.8       &  0.510    & 300   & 0.282     & d  \\
    \FRB{20180924B}        & 362    & 1297.5         &  21.1   &  41    & 0.864     & 1.76      &  0.234    & 18    & 0.244     & h  \\    
    \FRB{20181112A}        & 589    & 1297.5         &  19.3   &  40    & 0.864     & 2.1       &  0.312    & 28    & 0.481     & i  \\    
    \FRB{20190102C}        & 365    & 1271.5         &  14     &  57    & 0.864     & 1.7       &  0.229    & 16    & 0.230     & j  \\    
    \FRB{20190608B}        & 340    & 1271.5         &  16.1   &  37    & 1.728     & 6         &  0.362    & 28    & 0.224     & j  \\    
    \FRB{20190611B}        & 322    & 1271.5         &  9.5    &  58    & 1.728     & 2         &  0.481    & 10    & 0.182     & j  \\    
    \FRB{20190711A}        & 595    & 1271.5         &  23.8   &  57    & 1.728     & 6.5       &  0.079    & 36    & 0.470     & j  \\    
    \FRB{20190714A}        & 505    & 1271.5         &  10.7   &  39    & 1.728     & 2.9       &  0.458    & 13    & 0.396     & k  \\    
    \FRB{20191001A}        & 507    & 919.5          &  37.1   &  44    & 1.728     & 4.2       &  0.579    & 120   & 0.393     & l  \\    
    \FRB{20191228A}        & 298    & 1271.5         &  22.9   &  33    & 1.728     & 2.3       &  0.486    & 67    & 0.184     & m  \\    
    \FRB{20200430A}        & 380    & 864.5          &  13.85  &  27    & 1.728     & 6.5       &  0.475    & 35    & 0.278    & k  \\    
    \FRB{20200627A}        & 294    & 920.5          &  11.0   &  40    & 1.728     & 11        &  0.340     & 28    & 0.172       & o   \\
    \FRB{20200906A}        & 578    & 864.5          &  10.5   &  36    & 1.728     & 6         &  0.570    & 53    & 0.474     & m  \\   
    \FRB{20210117A}        & 730    & 1271.5         &  27.1   &  34    & 1.182     & 3.2       &  0.464    & 36    & 0.631     & n \\
    \FRB{20210214G}        & 398    & 1271.5         &  11.6   &  32    & 1.182     & 3.5       &  0.442    & 13    & 0.291    & o \\
    \FRB{20210320C}        & 384    & 864.5          &  15.3   &  42    & 1.182     & 5.4       &  1.146    & 59    & 0.266   & o \\
    \FRB{20210407E}        & 1785   & 1271.5         &  19.1   &  154   & 1.182     & 6.6       &  0.370     & 36    & 1.581      & o \\
    \FRB{20210807D}        & 252    & 920.5          &  47.1   &  121   & 1.182     & 10        &  0.602    & 100   & 0.034     & p  \\
    \FRB{20210809C}        & 652    & 920.5          &  16.8   &  190   & 1.182     & 14        &  0.190    & 45    & 0.392     & o,p  \\
    \FRB{20210912A}        & 1235   & 1271.5         &  31.7   &  31    & 1.182     & 5.5       &  0.475    & 70    & 1.146     & o,p  \\
    \FRB{20211127I}        & 235    & 1271.5         &  37.9   &  43    & 1.182     & 1.4       &  0.161    & 35    & 0.103     & p  \\
    \FRB{20211203C}        & 636    & 920.5          &  14.2   &  63    & 1.182     & 9.6       &  0.212    & 30    & 0.506     & o,q  \\
    \FRB{20211212A}        & 206    & 1632.5         &  12.8   &  27    & 1.182     & 2.7       &  0.772    & 131    & 0.089     & p  \\
    \FRB{20220105A}        & 583    & 1632.5         &  9.8    &  22    & 1.182     & 2.0       &  0.443    & 19    & 0.494     & o,q  \\
    \hline
    \end{tabular}\\
    ** These FRBs have $z_{\mathrm{DM}} < 0$, and are excluded from initial analysis. \\
    $\dag$ This FRB has S/N$_{\mathrm{cutoff}}=8$; all others are 9.5.
\end{table*}

\setcounter{table}{3}

\begin{table*}
    \centering
    \begin{tabular}{lrrrrrrrrrcl}
    \hline
    \multicolumn{1}{c}{Name} & \multicolumn{1}{c}{DM} & \multicolumn{1}{c}{$\nu_{c}$} & \multicolumn{1}{c}{S/N}  & \multicolumn{1}{c}{DM$_{\mathrm{Gal}}$} & \multicolumn{1}{c}{$\Delta t$} & \multicolumn{1}{c}{$w$} & \multicolumn{1}{c}{$\theta_{d}$} & \multicolumn{1}{c}{$F_{\nu}$} & \multicolumn{1}{c}{$z_{\mathrm{DM}}$} & \multicolumn{1}{l}{Reference} \\
    \multicolumn{1}{c}{} & \multicolumn{1}{c}{(pc cm$^{-3}$)} & \multicolumn{1}{c}{(MHz)} & \multicolumn{1}{c}{} & \multicolumn{1}{c}{} & \multicolumn{1}{c}{(pc cm$^{-3}$)} & \multicolumn{1}{c}{(ms)} & \multicolumn{1}{c}{(ms)} & \multicolumn{1}{c}{(\degree)} & \multicolumn{1}{c}{(Jy ms)} & \multicolumn{1}{c}{} & \multicolumn{1}{l}{}\\
    \hline
    \hline
     \FRB{20220501C}        & 450    & 863.5          &  16.1   &  31    & 1.182     & 6.5       &  0.516    & 32    & 0.347     & o,q  \\
    \FRB{20220531A}        & 727    & 1271.5         &  9.7    &  70    & 1.182     & 11.0      &  0.790    & 30    & 0.592     & o,q  \\
    \FRB{20220610A}        & 1458   & 1271.5         &  29.8   &  31    & 1.182     & 5.6       &  0.073    & 47    & 1.373     & r,s  \\
    \FRB{20220725A}        & 290    & 920.5          &  12.7   &  31    & 1.182     & 4.1       &  1.272    & 72   & 0.177     & o,q  \\
    \FRB{20220918A}        & 657    & 1271.5         &  26.4   &  41    & 1.182     & 7.1       &  0.457    & 55    & 0.550     & o,q  \\
    \FRB{20221106A}        & 344    & 1631.5         &  35.1   &  35    & 1.182     & 5.7       &  0.361    & 80    & 0.231     & o,q  \\
    \FRB{20230521A}         & 640.2  & 831.5          &  15.2   &  42  & 1.182     & 16.5      & 0.401     & 34    & 0.532   & o  \\
    \FRB{20230526A}         & 361.4  & 1271.5         &  22.1   &  50    & 1.182     & 4.7       & 0.383     & 34    & 0.233   & o  \\
    \FRB{20230708A}         & 411.5  & 920.5          &  31.5   &  50  & 1.182     & 23.6      & 0.657     & 111   & 0.286   & o  \\
    \FRB{20230718A}(**)         & 477    & 1271.5         &  10.9   &  396 & 1.182     & 3.5       & 0.317     & 14    & -0.02       & o  \\
    \FRB{20230731A}         & 701    & 1271.5         &  16.6   &  547 & 1.182     & 3.5       & 0.510      & 25    & 0.061       & o  \\
    \FRB{20230902A}         & 440    & 831.5          &  11.8   &  34  & 1.182     & 5.9       & 0.630      & 23    & 0.333       & o \\
    \FRB{20231006A}         & 509.7  & 863.5          &  15.2   &  68  & 1.182     & 8.3       & 0.534     & 25    & 0.371       & o  \\
    \FRB{20231226A}         & 329.9  & 863.5          &  36.7   &  38    & 1.182     & 11.8      & 0.739     & 78    & 0.213       & o  \\
    \hline
    \end{tabular}
 \caption{cont'd}
    ** These FRBs have $z_{\mathrm{DM}} < 0$, and are excluded from initial analysis.
\end{table*}

\begin{table*}
    \centering
     \caption
    {
        Derived properties of the \textit{\FullSample}\,for a fluence spectral index of $\alpha = 0.0$ and no source evolution. For those with $z_{\mathrm{DM}} < 0$ (marked with a $^*$: \FRB{20171020A}, \FRB{20180430A}, and \FRB{20230718A}), an assumed distance of 2\,Mpc is used. Columns are identical to those of Table~\ref{tab:LocalizedFRBDataTableExt}.
    \label{tab:UnlocalizedFRBDataTableExt}
    }
     \begin{tabular}{lrrrrrr}
    \hline
    \multicolumn{1}{c}{Name} & \multicolumn{1}{c}{$D_{L}$} & \multicolumn{1}{|c}{$E_{\nu}$} &
    \multicolumn{1}{c}{$z_{\max}$} & \multicolumn{1}{c}{$D_{L,\max}$}&
    \multicolumn{1}{c}{$V_{\mathrm{\max}}$} &
    \multicolumn{1}{c}{$\mathrm{V/}V_{\mathrm{\max}}$} \\
    \multicolumn{1}{c}{} & \multicolumn{1}{c}{(Gpc)} & \multicolumn{1}{|c}{(J Hz$^{-1}$)}  & \multicolumn{1}{c}{(Gpc)}& \multicolumn{1}{c}{(Gpc$^{3}$)} & \multicolumn{1}{c}{}\\
    \hline
    \hline
    \FRB{20170107A} & ${2.988}$ & $2.67 \cdot 10^{25}$ & ${ 0.696}$ & $1.75 \cdot 10^{7}$ & $2.25 \cdot 10^{7}$ & ${ 0.778}$  \\
\FRB{20170416A} & ${2.357}$ & $3.12 \cdot 10^{25}$ & ${ 0.590}$ & $1.03 \cdot 10^{7}$ & $1.39 \cdot 10^{7}$ & ${ 0.743}$ \\
\FRB{20170428A} & ${5.895}$ & $3.85 \cdot 10^{25}$ & ${ 0.970}$ & $4.25 \cdot 10^{7}$ & $4.27 \cdot 10^{7}$ & ${ 0.995}$ \\
\FRB{20170712A} & ${0.995}$ & $4.26 \cdot 10^{24}$ & ${ 0.256}$ & $1.41 \cdot 10^{6}$ & $1.74 \cdot 10^{6}$ & ${ 0.810}$ \\
\FRB{20170707A} & ${0.547}$ & $1.55 \cdot 10^{24}$ & ${ 0.119}$ & $2.15 \cdot 10^{5}$ & $2.15 \cdot 10^{5}$ & ${ 1.000}$  \\
\FRB{20170906A} & ${1.478}$ & $1.17 \cdot 10^{25}$ & ${ 0.410}$ & $4.02 \cdot 10^{6}$ & $6.13 \cdot 10^{6}$ & ${ 0.657}$ \\
\FRB{20171003A} & ${1.943}$ & $1.99 \cdot 10^{25}$ & ${ 0.486}$ & $7.26 \cdot 10^{6}$ & $9.71 \cdot 10^{6}$ & ${ 0.747}$ \\
\FRB{20171004A} & ${0.940}$ & $3.27 \cdot 10^{24}$ & ${ 0.215}$ & $1.06 \cdot 10^{6}$ & $1.11 \cdot 10^{6}$ & ${ 0.959}$  \\
\FRB{20171019A} & ${1.957}$ & $5.37 \cdot 10^{25}$ & ${ 0.736}$ & $8.56 \cdot 10^{6}$ & $2.29 \cdot 10^{7}$ & ${ 0.374}$  \\
\FRB{20171116A} & ${3.060}$ & $3.05 \cdot 10^{25}$ & ${ 0.675}$ & $1.70 \cdot 10^{7}$ & $2.02 \cdot 10^{7}$ & ${ 0.844}$ \\
\FRB{20171213A} & ${0.114}$ & $1.96 \cdot 10^{23}$ & ${ 0.054}$ & $5.29 \cdot 10^{3}$ & $2.17 \cdot 10^{4}$ & ${ 0.244}$ \\
\FRB{20171216A} & ${0.358}$ & $5.27 \cdot 10^{23}$ & ${ 0.099}$ & $9.96 \cdot 10^{4}$ & $1.25 \cdot 10^{5}$ & ${ 0.797}$ \\
\FRB{20180110A} & ${3.748}$ & $2.66 \cdot 10^{26}$ & ${ 1.996}$ & $3.48 \cdot 10^{7}$ & $1.38 \cdot 10^{8}$ & ${ 0.252}$ \\
\FRB{20180119A} & ${1.581}$ & $1.93 \cdot 10^{25}$ & ${ 0.436}$ & $4.67 \cdot 10^{6}$ & $7.10 \cdot 10^{6}$ & ${ 0.658}$ \\
\FRB{20180128B} & ${1.857}$ & $1.15 \cdot 10^{25}$ & ${ 0.413}$ & $5.60 \cdot 10^{6}$ & $6.16 \cdot 10^{6}$ & ${ 0.910}$  \\
\FRB{20180128A} & ${2.165}$ & $1.89 \cdot 10^{25}$ & ${ 0.464}$ & $7.97 \cdot 10^{6}$ & $8.61 \cdot 10^{6}$ & ${ 0.926}$ \\
\FRB{20180130A} & ${1.187}$ & $1.05 \cdot 10^{25}$ & ${ 0.288}$ & $2.05 \cdot 10^{6}$ & $2.35 \cdot 10^{6}$ & ${ 0.873}$  \\
\FRB{20180131A} & ${3.313}$ & $5.32 \cdot 10^{25}$ & ${ 0.841}$ & $2.13 \cdot 10^{7}$ & $3.12 \cdot 10^{7}$ & ${ 0.684}$\\
\FRB{20180212A} & ${0.196}$ & $4.05 \cdot 10^{23}$ & ${ 0.071}$ & $2.21 \cdot 10^{4}$ & $4.78 \cdot 10^{4}$ & ${ 0.462}$  \\
\FRB{20180315A} & ${1.653}$ & $1.06 \cdot 10^{25}$ & ${ 0.386}$ & $4.62 \cdot 10^{6}$ & $5.34 \cdot 10^{6}$ & ${ 0.866}$ \\
\FRB{20180324A} & ${1.581}$ & $1.25 \cdot 10^{25}$ & ${ 0.401}$ & $4.29 \cdot 10^{6}$ & $5.54 \cdot 10^{6}$ & ${ 0.774}$ \\
\FRB{20180417A} & ${2.125}$ & $1.36 \cdot 10^{25}$ & ${ 0.647}$ & $9.98 \cdot 10^{6}$ & $1.90 \cdot 10^{7}$ & ${ 0.524}$  \\
\FRB{20180515A} & ${1.294}$ & $5.84 \cdot 10^{24}$ & ${ 0.298}$ & $2.51 \cdot 10^{6}$ & $2.74 \cdot 10^{6}$ & ${ 0.915}$ \\
\FRB{20180525A} & ${1.517}$ & $4.93 \cdot 10^{25}$ & ${ 0.699}$ & $5.36 \cdot 10^{6}$ & $2.12 \cdot 10^{7}$ & ${ 0.253}$  \\
\FRB{20180924B} & ${1.287}$ & $2.32 \cdot 10^{24}$ & ${ 0.394}$ & $3.16 \cdot 10^{6}$ & $5.75 \cdot 10^{6}$ & ${ 0.551}$ \\
\FRB{20181112A} & ${2.817}$ & $1.18 \cdot 10^{25}$ & ${ 0.770}$ & $1.74 \cdot 10^{7}$ & $2.85 \cdot 10^{7}$ & ${ 0.611}$ \\
\FRB{20190102C} & ${1.205}$ & $1.81 \cdot 10^{24}$ & ${ 0.302}$ & $2.35 \cdot 10^{6}$ & $2.91 \cdot 10^{6}$ & ${ 0.809}$ \\
\FRB{20190608B} & ${1.174}$ & $3.03 \cdot 10^{24}$ & ${ 0.361}$ & $2.39 \cdot 10^{6}$ & $4.20 \cdot 10^{6}$ & ${ 0.570}$ \\
\FRB{20190611B} & ${0.934}$ & $7.35 \cdot 10^{23}$ & ${ 0.236}$ & $1.21 \cdot 10^{6}$ & $1.45 \cdot 10^{6}$ & ${ 0.838}$  \\
\FRB{20190711A} & ${2.740}$ & $1.46 \cdot 10^{25}$ & ${ 0.866}$ & $1.60 \cdot 10^{7}$ & $3.21 \cdot 10^{7}$ & ${ 0.499}$  \\
\FRB{20190714A} & ${2.240}$ & $3.92 \cdot 10^{24}$ & ${ 0.530}$ & $9.36 \cdot 10^{6}$ & $1.17 \cdot 10^{7}$ & ${ 0.802}$  \\
\FRB{20191001A} & ${2.220}$ & $3.57 \cdot 10^{25}$ & ${ 0.922}$ & $1.30 \cdot 10^{7}$ & $4.17 \cdot 10^{7}$ & ${ 0.311}$  \\
\FRB{20191228A} & ${0.940}$ & $4.98 \cdot 10^{24}$ & ${ 0.373}$ & $1.65 \cdot 10^{6}$ & $4.87 \cdot 10^{6}$ & ${ 0.338}$ \\
\FRB{20200430A} & ${1.491}$ & $5.60 \cdot 10^{24}$ & ${ 0.383}$ & $3.95 \cdot 10^{6}$ & $5.27 \cdot 10^{6}$ & ${ 0.749}$  \\
\FRB{20200627A} & ${0.873}$ & $1.83 \cdot 10^{24}$ & ${ 0.207}$ & $9.05 \cdot 10^{5}$ & $9.69 \cdot 10^{5}$ & ${ 0.934}$  \\
\FRB{20200906A} & ${2.768}$ & $2.18 \cdot 10^{25}$ & ${ 0.585}$ & $1.40 \cdot 10^{7}$ & $1.56 \cdot 10^{7}$ & ${ 0.900}$ \\
\FRB{20210117A} & ${3.890}$ & $2.39 \cdot 10^{25}$ & ${ 1.216}$ & $3.48 \cdot 10^{7}$ & $7.00 \cdot 10^{7}$ & ${ 0.497}$  \\
\FRB{20210214G} & ${1.575}$ & $2.27 \cdot 10^{24}$ & ${ 0.411}$ & $4.38 \cdot 10^{6}$ & $5.97 \cdot 10^{6}$ & ${ 0.734}$  \\
\FRB{20210320C} & ${1.421}$ & $8.72 \cdot 10^{24}$ & ${ 0.584}$ & $4.74 \cdot 10^{6}$ & $1.54 \cdot 10^{7}$ & ${ 0.308}$ \\
\FRB{20210407E} & ${11.853}$ & $8.86 \cdot 10^{25}$ & ${ 3.045}$ & $1.74 \cdot 10^{8}$ &$2.52 \cdot 10^{8}$ & ${ 0.693}$ \\
\FRB{20210807D} & ${0.163}$ & $2.96 \cdot 10^{23}$ & ${ 0.100}$ & $1.65 \cdot 10^{4}$ & $1.27 \cdot 10^{5}$ & ${ 0.130}$  \\
\FRB{20210809C} & ${2.213}$ & $1.33 \cdot 10^{25}$ & ${ 0.577}$ & $9.33 \cdot 10^{6}$ & $1.33 \cdot 10^{7}$ & ${ 0.699}$  \\
\FRB{20210912A} & ${8.011}$ & $1.14 \cdot 10^{26}$ & ${ 3.359}$ & $1.20 \cdot 10^{8}$ & $2.76 \cdot 10^{8}$ & ${ 0.436}$  \\
\FRB{20211127I} & ${0.506}$ & $8.73 \cdot 10^{23}$ & ${ 0.220}$ & $3.48 \cdot 10^{5}$ & $1.21 \cdot 10^{6}$ & ${ 0.286}$  \\
\FRB{20211203C} & ${2.988}$ & $1.38 \cdot 10^{25}$ & ${ 0.667}$ & $1.62 \cdot 10^{7}$ & $1.93 \cdot 10^{7}$ & ${ 0.836}$  \\
\FRB{20211212A} & ${0.432}$ & $2.44 \cdot 10^{24}$ & ${ 0.285}$ & $2.58 \cdot 10^{5}$ & $2.29 \cdot 10^{6}$ & ${ 0.113}$  \\
\FRB{20220105A} & ${2.903}$ & $8.39 \cdot 10^{24}$ & ${ 0.721}$ & $1.64 \cdot 10^{7}$ & $2.28 \cdot 10^{7}$ & ${ 0.718}$  \\
\FRB{20220501C} & ${1.923}$ & $7.65 \cdot 10^{24}$ & ${ 0.517}$ & $7.51 \cdot 10^{6}$ & $1.14 \cdot 10^{7}$ & ${ 0.660}$  \\
\FRB{20220531A} & ${3.600}$ & $1.79 \cdot 10^{25}$ & ${ 1.387}$ & $2.86 \cdot 10^{7}$ & $7.41 \cdot 10^{7}$ & ${ 0.385}$ \\
\FRB{20220610A} & ${9.979}$ & $9.70 \cdot 10^{25}$ & ${ 2.939}$ & $1.49 \cdot 10^{8}$ & $2.45 \cdot 10^{8}$ & ${ 0.609}$ \\
\FRB{20220725A} & ${0.903}$ & $5.00 \cdot 10^{24}$ & ${ 0.447}$ & $1.68 \cdot 10^{6}$ & $7.96 \cdot 10^{6}$ & ${ 0.211}$  \\
    \hline
    \end{tabular}
\end{table*}

\setcounter{table}{4}
\begin{table*}
    \centering
     \caption
    {
        (cont'd).
    }
     \begin{tabular}{lrrrrrr}
    \hline
    \multicolumn{1}{c}{Name} & \multicolumn{1}{c}{$D_{L}$} & \multicolumn{1}{|c}{$E_{\nu}$} &
    \multicolumn{1}{c}{$z_{\max}$} & \multicolumn{1}{c}{$D_{L,\max}$}&
    \multicolumn{1}{c}{$V_{\mathrm{\max}}$} &
    \multicolumn{1}{c}{$\mathrm{V/}V_{\mathrm{\max}}$} \\
    \multicolumn{1}{c}{} & \multicolumn{1}{c}{(Gpc)} & \multicolumn{1}{|c}{(J Hz$^{-1}$)}  & \multicolumn{1}{c}{(Gpc)}& \multicolumn{1}{c}{(Gpc$^{3}$)} & \multicolumn{1}{c}{}\\
    \hline
    \hline
\FRB{20220918A} & ${3.299}$ & $2.91 \cdot 10^{25}$ & ${ 1.416}$ & $2.59 \cdot 10^{7}$ & $8.15 \cdot 10^{7}$ & ${ 0.318}$ \\
\FRB{20221106A} & ${1.212}$ & $9.12 \cdot 10^{24}$ & ${ 0.670}$ & $3.28 \cdot 10^{6}$ & $1.80 \cdot 10^{7}$ & ${ 0.183}$ \\
\FRB{20230521A} & ${3.171}$ & $1.70 \cdot 10^{25}$ & ${ 0.779}$ & $1.91 \cdot 10^{7}$ & $2.64 \cdot 10^{7}$ & ${ 0.724}$  \\
\FRB{20230526A} & ${1.227}$ & $3.95 \cdot 10^{24}$ & ${ 0.455}$ & $2.95 \cdot 10^{6}$ & $7.56 \cdot 10^{6}$ & ${ 0.390}$  \\
\FRB{20230708A} & ${1.545}$ & $1.88 \cdot 10^{25}$ & ${ 0.805}$ & $5.64 \cdot 10^{6}$ & $2.69 \cdot 10^{7}$ & ${ 0.210}$ \\
\FRB{20230731A} & ${0.290}$ & $2.22 \cdot 10^{23}$ & ${ 0.108}$ & $6.83 \cdot 10^{4}$ & $1.61 \cdot 10^{5}$ & ${ 0.424}$  \\
\FRB{20230902A} & ${1.835}$ & $5.11 \cdot 10^{24}$ & ${ 0.441}$ & $6.34 \cdot 10^{6}$ & $7.89 \cdot 10^{6}$ & ${ 0.803}$ \\
\FRB{20231006A} & ${2.079}$ & $6.73 \cdot 10^{24}$ & ${ 0.551}$ & $8.65 \cdot 10^{6}$ & $1.29 \cdot 10^{7}$ & ${ 0.670}$ \\
\FRB{20231226A} & ${1.105}$ & $7.63 \cdot 10^{24}$ & ${ 0.602}$ & $2.71 \cdot 10^{6}$ & $1.51 \cdot 10^{7}$ & ${ 0.179}$ \\
    \hline
\FRB{20171020A}$^*$ & 0.00103 & $2.68 \cdot 10^{22}$ & 0.00050    & 0.00392             & 0.0177              &  0.222 \\
\FRB{20180430A}$^*$ & 0.00103 & $2.2 \cdot 10^{22}$  & 0.00048    & 0.00384             & 0.0154              &  0.249 \\
\FRB{20230718A}$^*$ & 0.00103 & $1.80 \cdot 10^{18}$ & 0.00028    & 0.00258             & 0.00292             &  0.882 \\
    \hline
    \end{tabular}
\end{table*}

\section{Conclusions}
\label{sec:Conclusion}

We have shown how to apply the \VonVmax\ method of \citet{Schmidt_1968} to a population of transient sources and applied this to fast radio bursts. We find that the current sample of FRBs detected by ASKAP/CRAFT is insufficient to distinguish between competing evolutionary and spectral models, with the greatest departure from uniformity in the \VonVmax\ distribution being due to a dearth of very high signal-to-noise FRBs.

Using both FRBs with known redshift, $z_{\rm loc}$, and a larger sample of FRBs with $z_{\rm DM}$ estimated from the Macquart relation, we plot the FRB energy distribution in the range $10^{23}$--$10^{26}$\,J\,Hz$^{-1}$. We find it to be fairly consistent ($p=0.09$) with a power-law with differential slope $\gamma=-1.96 \pm 0.15$ using $z_{\rm loc}$. Above this energy, we find some evidence of a downturn consistent with a Schechter function with $E_{\rm max} = 6.3 \, \cdot 10^{25}$ when using $z_{\rm DM}$. We have also identified several systematic effects in the analysis, and shown how to take these into account. In particular, the difficulty of identifying high-$z$ host galaxies limits our knowledge of the tip of the FRB energy distribution, as it is unclear if the downturn in the energy distribution seen in the $z_{\rm DM}$ result is physical, or an artefact of smearing in the Macquart relation.

In the near future, FRB surveys will detect too many bursts to follow up their host galaxies individually with 8\,m-class telescope time \citep[e.g.\ CHORD; ][]{2019clrp.2020...28V}. Low-DM, near-Universe host galaxies can likely be identified in existing or impending (e.g., LSST) optical surveys without further follow-up, allowing an unbiased sample of the $E_\nu < 10^{23}$\,J\,Hz$^{-1}$ region to be formed. Moreover, we find that the use of $z_{\rm DM}$ compared to $z_{\rm loc}$ does not significantly affect the luminosity function in the range $10^{23}$--$10^{26}$\,J\,Hz$^{-1}$. We therefore recommend that optical follow-up time be focused on identifying high-DM/$z_{\rm loc}$ FRBs, to allow the high-end of the FRB luminosity function to be studied.

\clearpage
\begin{acknowledgement}
This scientific work uses data obtained from Inyarrimanha Ilgari Bundara / the Murchison Radio-astronomy Observatory. We acknowledge the Wajarri Yamaji People as the Traditional Owners and native title holders of the Observatory site. CSIRO’s ASKAP radio telescope is part of the Australia Telescope National Facility (https://ror.org/05qajvd42). Operation of ASKAP is funded by the Australian Government with support from the National Collaborative Research Infrastructure Strategy. ASKAP uses the resources of the Pawsey Supercomputing Research Centre. Establishment of ASKAP, Inyarrimanha Ilgari Bundara, the CSIRO Murchison Radio-astronomy Observatory and the Pawsey Supercomputing Research Centre are initiatives of the Australian Government, with support from the Government of Western Australia and the Science and Industry Endowment Fund.

\end{acknowledgement}

\paragraph{Funding Statement}

W.A. acknowledges the contribution of an Australian Government Research Training Program Scholarship in support of this research. C.W.J.\ and M.G.\ acknowledge support by the Australian Government through the Australian Research Council's Discovery Projects funding scheme (project DP\,210102103). A.T.D.\ and R.M.S.\ acknowledge support through ARC Discovery Project DP\,220102305.  R.M.S.\ acknowledges support through ARC Discovery Project DP 220102305. 
This research was supported by the Australian Research Council Centre of Excellence for All Sky Astrophysics in 3 Dimensions (ASTRO 3D) through project no.\,CE 170100013. 
A.C.G.\ and the Fong Group at Northwestern acknowledges support by the National Science Foundation under grant Nos. AST-1909358, AST-2308182 and CAREER grant No. AST-2047919. A.C.G.\ acknowledges support from NSF grants AST-1911140, AST-1910471 and AST-2206490 as a member of the Fast and Fortunate for FRB Follow-up team.

\paragraph{Competing Interests}

None.

\paragraph{Data Availability Statement}

Data underlying this article are available within this article.
Code to generate \VonVmax\ and luminosity functions is contained in the { \sc FRB }  repository \citep{zdm}.
Digitised versions of the figures are available upon reasonable request to the authors.


\printendnotes

\printbibliography

\appendix
\section{Investigation of Spectral Dependence and Source Evolution}
\label{sec:spectral_source}

For simplicity, in the main body of this work, we treated the case of no spectral dependence and no cosmological source evolution. Here, we show that with current data, the \VonVmax\ test cannot determine whether either effect is present, and show the resulting systematic effects on the luminosity function.

The spectral dependence of FRBs is still uncertain. \citet{Macquartetal2019} used ASKAP FRBs to determine a spectral dependence of $F_\nu \propto \nu^\alpha$, with $\alpha = -1.5_{-0.3}^{+0.2}$, though as noted by \citet{James_etal_2021_1}, selection biases due to FRBs being narrow-band might imply the true dependence is $\alpha = -0.65$. The apparent rate of FRBs measured by CHIME appears to be frequency-independent; however as noted by the authors, this does not account for selection biases \citep{CHIME_catalog1_2021}. Population modelling by \citet{James2022H0} and \citet{Shin2022} find some evidence for increased spectral strength at lower frequencies, however constraints are very weak, as are those from studies of the frequency-dependent detection rate measured by ASKAP. We therefore consider both $\alpha = 0$ and $\alpha = -1.5$ in this investigation.


The source evolution function, $\psi$, weights the physical volume, $V$, to produce an effective volume, $V^\prime$. If the source density in the Universe varies with redshift, then only the distribution of $V^\prime/V_{\rm max}^{\prime}$ will be uniform between $0$ and $1$. The source evolution function $\psi(z)$ is inserted into the integrals over redshift, viz., eq.~(\ref{eqn:Vmax2}) and eq.~(\ref{eqn:Vol2}) to calculate $V_{\rm max}^\prime$ and $V^\prime$ respectively. Since $V^\prime = V$ only in the case that the FRB population does not cosmologically evolve --- a situation which we do not deem likely --- we henceforth drop the ${}^\prime$ notation so that both $V$ and $V_{\rm max}$ are implicitly understood to be weighted by $\psi(z)$.

We consider source evolution by scaling $V$ to some power of the star formation rate as parameterised by \citet{MadauDickinson_SFR},
\begin{eqnarray}
{\rm SFR}(z) & \propto & \frac{(1+z)^{2.7}}{1 + \left(\frac{1+z}{2.9}\right)^{5.6}}, \label{eq:sfr}\\ 
\psi(z) & = & \left( {\rm SFR}(z)\right)^{n_{\rm SFR}}. \label{eq:sfr_n}
\end{eqnarray}
Given that the majority of this sample represents the $z<0.5$ Universe, where the denominator of eq.~(\ref{eq:sfr}) changes by at most 2.5\%, this scaling is almost equivalent to a scaling of $\psi(z) = (1+z)^{2.7 n_{\rm SFR}}$.

\begin{figure}
    \centering
    \includegraphics[width = \columnwidth]{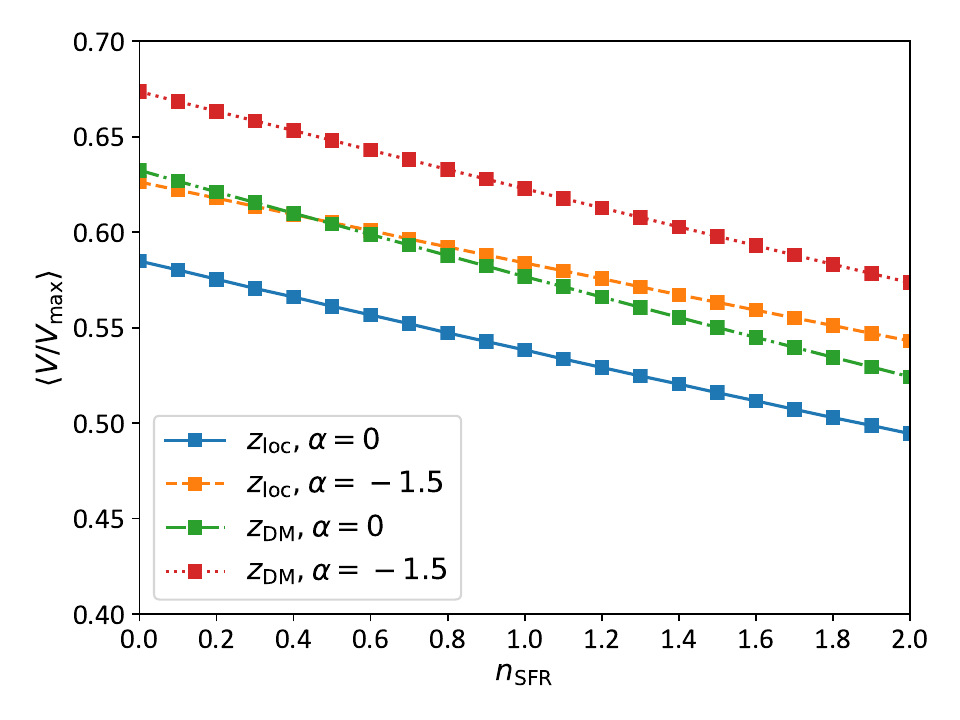}
    \caption
    { Calculated values of $\left< V/V_{\rm max} \right>$, considering two values of the spectral index $\alpha = \left\{0,-1.5\right\}$, for both the \LocalizedSample\ (using $z_{\rm loc}$) and the \FullSample\ (using $z_{\rm DM}$), as a function of the star formation rate scaling parameter $n_{\rm SFR}$.
    }
    \label{fig:meanVonVmax}
\end{figure}

\begin{figure}
    \centering
    \includegraphics[width = \columnwidth]{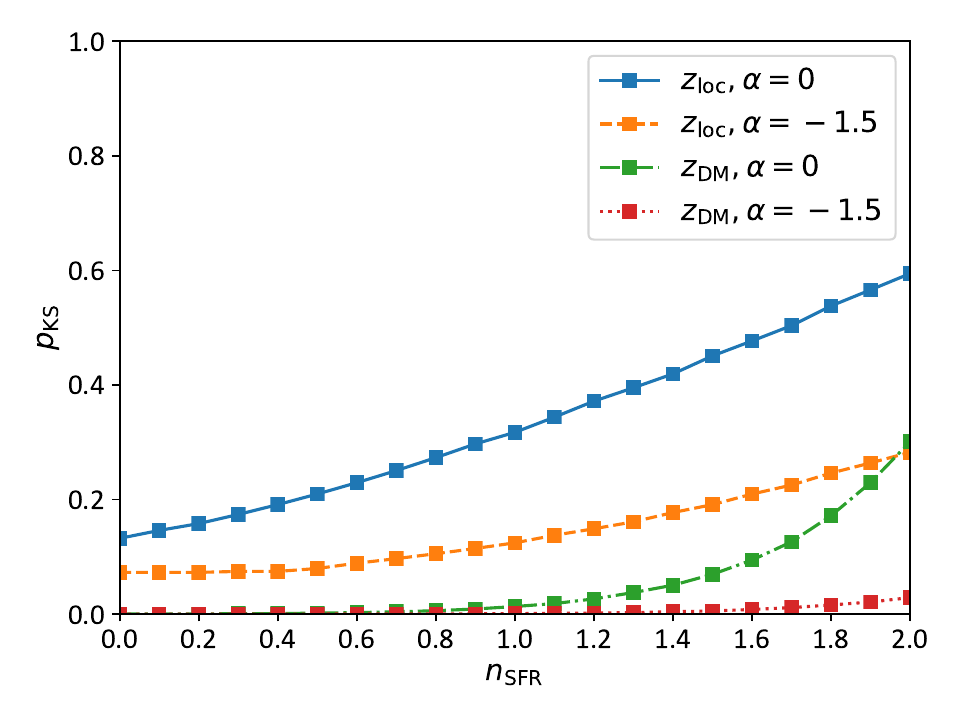}
    \caption
    {P-values resulting from the KS-test for uniformity in $V/V_{\rm max}$, considering two values of the spectral index $\alpha = \left\{0,-1.5\right\}$, for both the \LocalizedSample\ (using $z_{\rm loc}$) and the \FullSample\ (using $z_{\rm DM}$), as a function of the star formation rate scaling parameter $n_{\rm SFR}$.
    }
    \label{fig:ksresults}
\end{figure}

\begin{figure}
    \centering
        \includegraphics[width=\columnwidth]{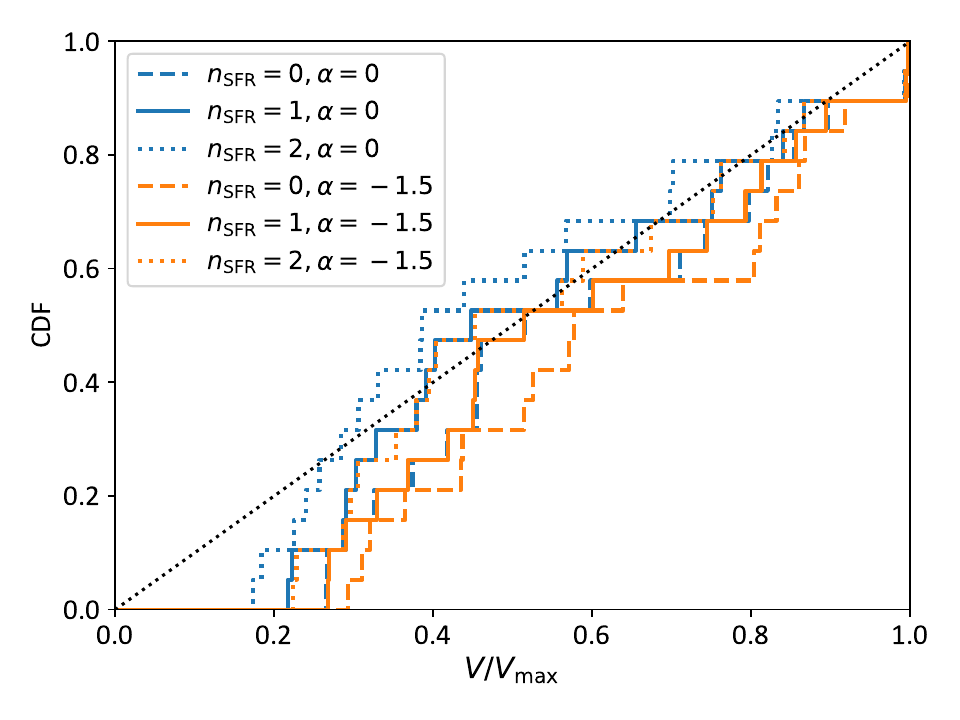}
    \caption
    {Cumulative histograms of $V/V_{\rm max}$ for six combinations of $\alpha$ and $n_{\rm SFR}$ for the \LocalizedSample, compared to the expectation (black dotted line). Note that the $n_{\rm SFR} = 0, \alpha = 0$ \text{and} $n_{\rm SFR} = 1, \alpha = -1.5$ plots almost overlap, as do the $n_{\rm SFR} = 1, \alpha = 0$ and $n_{\rm SFR} = 2, \alpha = -1.5$ plots.
    }
    \label{fig:cumulative_hists}
\end{figure}

Figure~\ref{fig:meanVonVmax} plots the $\langle V/V_{\mathrm{max}}\rangle$ values for both the \LocalizedSample\,and \FullSample\, along with their 95\% confidence intervals, determined using the bootstrap method described in Appendix \ref{sec:errors}. 
To check for population uniformity, we further conduct a Kolmogorov-Smirnoff (K-S)-test with respect to a uniform distribution for both samples. The resulting p-values are shown in Figure~\ref{fig:ksresults}.

\begin{figure}
    \centering
        \includegraphics[width=\columnwidth]{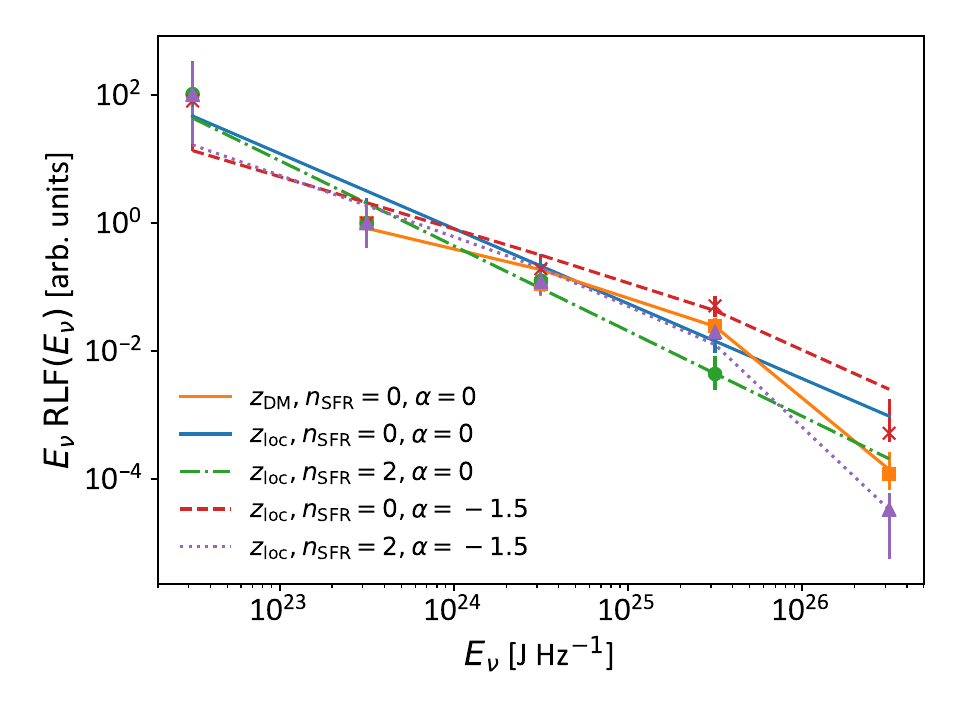}
    \caption
    {Radio luminosity functions (RLFs) calculated from  the \LocalizedSample\ (using $z_{\rm loc}$) for combinations of $\alpha = \left\{0,-1.5\right\}$ and $n_{\rm SFR} = \left\{0,2\right\}$, and from the \FullSample\ (using $z_{\rm DM}$) for $\alpha=0,n_{\rm SFR}=0$. The (arbitrary) normalisation is fixed to unity at the $10^{23}$--$10^{24}$ bin. The best-fit Schechter functions for each sample are depicted for reference purposes.
    }
    \label{fig:appRLFPlots}
\end{figure}

\subsection{Uniformity of $V/V_{\rm max}$}

Varying both $\alpha$ and $n_{\rm SFR}$ produces our results on the uniformity of $V/V_{\rm max}$ shown in Figures~\ref{fig:meanVonVmax} to \ref{fig:cumulative_hists}. Requiring only that $\left< V / V_{\rm max} \right>=0.5$ favours a strongly evolving FRB population, with $n_{\rm SFR}=1.7$ for $\alpha=0$, and $n_{\rm SFR}>2$ for $\alpha=-1.5$. Both the \LocalizedSample\ and \FullSample\ yield almost identical values of $\left< V / V_{\rm max} \right>$. The p-values of the KS-statistics shown in Figure~\ref{fig:ksresults} confirm this, however at the $2 \sigma$ level (p < $0.05$), no value of $n_{\rm SFR}$ is excluded for the \LocalizedSample, while the \FullSample\ shows stronger evidence against uniformity for low $n_{\rm SFR}$.

The driver of these results, as shown in Figure~\ref{fig:cumulative_hists}, is the lack of events with very low \VonVmax\ --- equivalently, a lack of very high S/N events. Indeed, none of the cumulative \VonVmax\ distributions give a very good fit to uniformity. We have considered in \shannon\ whether or not this effect could be due to instrumental bias, and concluded that high S/N events would still be detectable in adjacent beams even if a primary beam was saturated. We therefore conclude that the lack of low $V/V_{\rm max}$ events is probably a statistical under-fluctuation, and that uniformity in \VonVmaxbar\ does not currently discriminate between different values of $n_{\rm SFR}$ and $\alpha$.
Figure~\ref{fig:cumulative_hists} also illustrates the degeneracy between $n_{\rm SFR}$ and $\alpha$: a steeper spectral index, and hence k-correction, allows for a more strongly evolving source population, as noted by \citet{James_etal_2021_1}.

Our inability to distinguish between plausible values of $\alpha$ and $n_{\rm SFR}$ results in a difference in the behaviour of the luminosity functions at high energies, as shown in Figure~\ref{fig:appRLFPlots}. No spectral evolution ($\alpha=0$) predicts distributions consistent with a pure power-law, while $\alpha=-1.5$ produces a high-energy downturn consistent with the Schechter function. The effect of increasing $n_{\rm SFR}$ is primarily to produce a stronger downturn (lower $E_{\rm max}$), though this is only evident for $\alpha = -1.5$.

The uncertainty in the luminosity function, due to our inability to determine the population evolution or the spectral dependence with \VonVmax, is comparable to the systematic errors identified in the \FullSample\, and \LocalizedSample\, discussed in Section~\ref{sec:biases}. However, this method could be used to constrain these parameters in a future analysis.
\section{Using $V/V_{\mathrm{max}}$ with Non-uniform Sensitivity}
\label{sec:VonVMaxNonUniformSensitivity}

The original formulation of the $V/V_{\mathrm{max}}$-test by \citet{Schmidt_1968} was provided in the context of optical and radio quasar surveys with well-defined luminosity thresholds, $S_{\mathrm{cutoff}}$, and survey areas, $\Omega$. This allowed for conceptually easy definitions of survey volumes $V$ and $V_{\mathrm{max}}$ for a given cosmology. For transient sources such as FRBs however, the definition of these quantities becomes less obvious. Here we show how to construct $V$ and $V_{\mathrm{max}}$ in the case of spatial- and time-varying sensitivity.

\subsection{Spatially Varying Sensitivity}
\label{sec:varyingSensitivy}

FRBs are transients, and as such they will be observed at a particular part of a telescope's beam, with sensitivity, $B$, with respect to the beam centre (where $B=1$). Unlike steady sources, where multiple pointings can, to a large extent, correct for sources viewed far from the beam centre, $S_{\mathrm{cutoff}}$ --- or in our formulation, $F_{\nu,{\mathrm{cutoff}}}$ --- varies over solid angle, hence from event-to-event. While this approach generalises to any spatially varying sensitivity, we consider only the beamshape $B$, where $F_{\nu,{\mathrm{cutoff}}} \propto B^{-1}$, hereinafter.

One approach (a differential method) to deal with this is to consider only an infinitesimal solid angle, $d\Omega$, about the point of detection. In this case, the fluence cutoff, $F_{\nu,{\mathrm{cutoff}}}$, is well-defined since the beam sensitivity is locally constant. Each and every detection therefore becomes its own survey over an infinitesimal solid angle $d\Omega$, resulting in infinitesimally small $V \to dV$ and $V_{\mathrm{max}} \to dV_{\mathrm{max}}$. In such a case, the absolute values of $dV$ and $dV_{\mathrm{max}}$ have little meaning, preventing the total source density from being derived; their ratio, however, is well-defined and preserves the properties of the $V/V_{\mathrm{max}}$-test.

\begin{figure}
    \centering
    \includegraphics[width=\columnwidth]{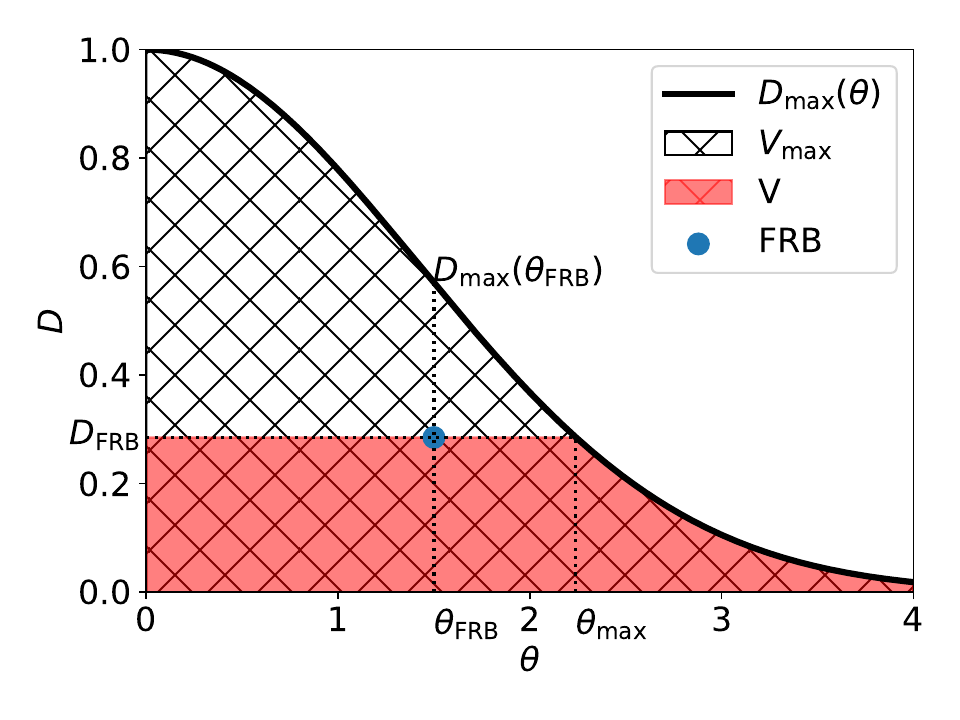}
    \caption{Illustration of the volumes $V$ and $V_{\mathrm{max}}$ for an FRB detected at distance $D_{\mathrm{FRB}}$ at position $\theta_{\mathrm{FRB}}$ away from the beam centre.}
    \label{fig:volumes}
\end{figure}

An alternative approach (an integral method) is to consider the total volume viewed by the telescope beam and the regions over which the FRB could have been detected. This situation is illustrated in Figure~\ref{fig:volumes}. Suppose an FRB is detected at position $\theta_{\mathrm{FRB}}$ away from the beam centre, and came from a distance $D_{\mathrm{FRB}}$; where it could have been detected out to a distance $D_{\mathrm{max}}(\theta_{\mathrm{FRB}})$ at that position in beam.

Since the beam sensitivity varies with position on the sky, the event at distance $D_{\mathrm{FRB}}$ would have been detectable at any point in the beam between the beam centre ($\theta=0$) and some maximum angle $\theta_{\mathrm{max}}$. However, it could have been detected at a maximum distance $D_{\mathrm{max}}(\theta)$ that varies with beam angle $\theta$. Therefore, the total volume $V_{\mathrm{max}}$ in which the FRB could have been detected is the region contained beneath the $D_{\mathrm{max}}(\theta)$ curve, while the volume $V$ in which it was detected is the same region, albeit limited by the actual distance to the event, $D_{\mathrm{FRB}}$.

It is interesting to compare the results of the integral method with that of the differential method, where $V$ and $V_{\mathrm{max}}$ depend only on the values $D_{\mathrm{FRB}}$ and $D_{\mathrm{max}}(\theta_{\mathrm{FRB}})$ at the point $\theta_{\mathrm{FRB}}$ --- the point at which the FRB was detected. Clearly, for any given event, the value $V/V_{\mathrm{max}}$ will be different between the two methods. Yet, statistically, they give identical results.

We have tested the differential and integral methods using a simple simulation of FRBs distributed in a Euclidean space viewed by a 2-dimensional Gaussian beamshape. We generated a sample of $10^6$ FRBs uniformly in the sensitive volume, and calculated $V/V_{\mathrm{max}}$ for each simulated FRB using both methods. In both cases a uniform distribution of $V/V_{\mathrm{max}}$ over the range $\left[0, 1\right]$ was obtained within statistical errors.

\subsection{Time-varying Sensitivity}
\label{sec:timeVaryingSensitivy}

Time-variation of survey sensitivity, $F_{\nu,{\mathrm{cutoff}}}$, is no different to variation over a beam pattern --- it is just another dimension. Analogously, a transients survey is characterised not just by the survey area, $\Delta \Omega$, and threshold, $F_{\nu,{\mathrm{cutoff}}}$, it is also characterised by its duration, $T_{\mathrm{obs}}$. Likewise, the instantaneous volume element of the Universe in which transients occur is $d\Omega dzd\tau$, where proper time, $d\tau$, is simply another dimension of the volume.

Furthermore, the sensitivity of FRB surveys can also vary with time, either on rapid timescales (e.g., due to RFI) or on slow timescales (e.g., due to varying telescope configurations). The latter is a particular problem for commensal observations.

The differential and integral methods discussed above therefore apply identically to the time dimension. The differential method requires knowing the survey sensitivity only at the time of detection, whereas the integral method requires knowing the survey sensitivity for the entire duration of the survey, and integrating the volumes $V(t)$ and $V_{\mathrm{max}}(t)$ over survey time, $t$. 

\subsection{Application to the Current Work}
\label{sec:apptotheCurrentWork}

Holographic observations have allowed accurate measurements of ASKAP's beamshape \citep{Jamesetal2019} to be made, allowing the integral method to be used to account for ASKAP's spatial variation in sensitivity over $\Delta \Omega$. However, a proper accounting of changing detector conditions with time makes the integral method too complex to deal with this dimension; we therefore use the differential method in the time domain for our analysis, by taking the survey conditions at the instant at which each FRB has been detected.

\section{Error calculations for the luminosity function}
\label{sec:errors}

The luminosity histogram is built by summing the inverse values of \VMax. Treating this process as a weighted sum produces an error corresponding to
\begin{eqnarray}
\sigma_v & = \ & \sqrt{\sum_{i=1}^{N} \frac{1}{(V_{\rm max}^i)^2}},
\end{eqnarray}
for $N$ FRBs in a histogram bin. Equivalently, we can use resampling --- replacing each FRB with $M$ copies of itself, where $M$ is an integer sampled from a Poissonian distribution of mean unity --- to estimate the error. These methods produce statistically identical estimates of $\sigma_v$. However, both formally treat the problem of ``if the observation is the truth, what is the plausible range of alternate observations?'' rather than the inverse ``what range of plausible truths could reproduce this observation?''. While the latter formulation is formally correct, for many statistical problems, the difference between these two statements is small. Here, however, different values of beam efficiency $B$, and sparse histogram binning, lead to \VMax\ varying by up to a factor of $300$ within a given bin, so that individual samples dominate, and the effective sample size approaches unity. This then leads to the uncertainty in that bin being comparable to the value in the bin itself, which is a clear miscalculation of the error.

To estimate the error in each luminosity function bin therefore, we use the bootstrap resampling method above, but vary the expected mean of the Poissonian distribution by a factor $k$. We generate lower (upper) limits on each bin by finding the smallest (largest) factor $k$ such that $0.5 (1-0.6827) = 15.865\%$ of resampled values are greater than (less than) the measured value. The lower (upper) bound then becomes that bin value multiplied by $k$. For this purpose, we use $10^{4}$ resamplings per bin.

\end{document}